\documentclass[aps,twocolumn,showpacs,preprintnumbers,amsmath,amssymb,
epsf,superscriptaddress,groupedaddress, tightenlines]{revtex4}

\usepackage{graphicx}
\usepackage{amsmath}
\usepackage{bm}
\usepackage{color}
\newcommand{\nn}{\nonumber}
\newcommand{\vslash}{v\hspace*{-5.5pt}\slash}

\newcommand{\Bslash}{{\cal B}\hspace*{-5.5pt}\slash}

\def\nslash{n\!\!\!\slash}
\def\bnslash{\bar n\!\!\!\slash}
\def\bn{\bar n}
\newcommand{\ri}[1]{{\color{red} #1}}

\def\OMIT#1{}

\begin{document}
\preprint{MIT-CTP 3625}
\preprint{UCSD/PTH 05-05}

\title{
Factorization in $B \to K \pi \ell^+ \ell^-$ decays}


\author{Benjam\'\i{}n Grinstein}
\affiliation{Department of Physics, University of California at San Diego,
  La Jolla, CA 92093}

\author{Dan Pirjol}
\affiliation{Center for Theoretical Physics, Massachusetts Institute for
  Technology, Cambridge, MA 02139}


\date{\today}

\begin{abstract}
We derive factorization relations
for the transverse helicity amplitudes in $B\to K\pi \ell^+ \ell^-$ at
leading order in $\Lambda/m_b$, in the kinematical region with an energetic
kaon and a soft pion. We identify and compute a new contribution of leading
order in $\Lambda/m_b$
to the $B\to K\pi\ell^+\ell^-$ amplitude which is not present in the
one-body decay $B\to K^* \ell^+\ell^-$. As an application we study the
forward-backward asymmetry (FBA) of the lepton momentum angular distribution 
in $B\to K\pi \ell^+ \ell^-$ decays away from the $K^*$ resonance.
The FBA in these decays has a zero at 
$q_0^2 = q_0^2(M_{K\pi})$, which can be used, in principle,  for determining 
the Wilson coefficients $C_{7,9}$ and testing the Standard Model. We point out 
that the slope of the $q_0^2(M_{K\pi}^2)$ curve contains the same information 
about the Wilson coefficients as the location
of the zero, but is less sensitive to unknown nonperturbative dynamics. We estimate
the location of the zero at leading order in factorization, and using a resonant
model for the $B\to K\pi\ell^+\ell^-$ nonfactorizable amplitude.
\end{abstract}

\pacs{12.39.Fe, 14.20.-c, 13.60.-r}


\maketitle


\section{Introduction}

The rare electroweak penguin decays $b\to s\gamma$ and $b\to s
\ell^+ \ell^-$ are sensitive probes of the flavor structure of the
Standard Model, and provide a promising testing ground for the study
of new physics effects (see Ref.~\cite{Ali} for a recent review of the
experimental situation).
Several clean tests have been proposed in these decays, which are sensitive 
to the chiral structure of the quark couplings in the Standard Model. 
Examples of such tests involve measuring 
the photon polarization in $b\to s\gamma$
and the zero of the forward-backward asymmetry in $b\to s\ell^+\ell^-$ \cite{fb}.

Our understanding of these decays has advanced considerably over the
past few years, through the derivation of factorization relations for
exclusive $B\to K^{(*)} \ell^+\ell^-$ and $B\to K^* \gamma$  decays at large recoil. 
First derived at lowest order in perturbation theory \cite{befe,fbcor1}, 
these factorization theorems were proved to all orders in $\alpha_s$ 
\cite{bpsff,ps1,Beneke:2003pa,hillff,CK,pol,BHN}
using the soft-collinear effective theory \cite{scet,scet1}.

In this paper we introduce a new factorization relation for the multibody 
rare decays
$B\to K\pi\ell^+\ell^-$ in the kinematical region with a soft pion and an 
energetic kaon, at leading order in $\Lambda/m_b$. The schematic form of the 
factorization relation is given below in Eq.~(\ref{factmulti}). This extends the
application of factorization to final states in $B\to X_s \ell^+\ell^-$ 
containing a few hadrons, with small total invariant mass. 

A particularly clean test for new physics effects in these decays is based 
on the forward-backward asymmetry of the (charged) lepton momentum in 
$B\to K^* \ell^+ \ell^-$ 
with respect to the decay axis $q = p_{\ell^+} + p_{\ell^-}$. 
This is defined as
\begin{multline}\label{AFBdef}
 A_{FB}(q^2) = 
 \frac1{d\Gamma(q^2)/dq^2}\Bigg[\int_{0}^{1} \mbox{d} \cos\theta_+ 
 \frac{d\Gamma(q^2,\theta_+)}{dq^2 d\cos\theta_+}\\
- \int_{-1}^0 \mbox{d} \cos\theta_+ \frac{d\Gamma(q^2,\theta_+)}{dq^2 d\cos\theta_+}\Bigg]
\end{multline}
where $\theta_+$ is the angle between $\vec p_{\ell^+}$ and $\vec q$ in the
rest frame of the lepton pair.

As pointed out in \cite{fb,ABHH}, due to certain form factor relations at large 
recoil \cite{ffrel,befe}, this asymmetry has a zero at $q_0^2$
which depends mostly on Wilson coefficients in the weak Hamiltonian
with little hadronic uncertainty. 
The position of the zero was computed in \cite{fbcor1,fbcor2,fbcor3}, using
the complete leading order factorization formula.
The most updated result, including isospin violation effects, is \cite{fbcor3}
\begin{eqnarray}\label{1}
q_0^2 &=& - 2m_B m_b \frac{\mbox{Re}(C_7^{\rm eff}(q_0^2))}{\mbox{Re }(C_9^{\rm eff}(q_0^2))}
(1+\tau(\alpha_s)) + 
\delta_{\rm fact} \\ 
& &= \left\{
\begin{array}{cc}
4.15\pm 0.27 \mbox{ GeV}^2 & (B^+\to K^{*+}) \\
4.36^{+0.33}_{-0.31} \mbox{ GeV}^2 & (B^0\to K^{*0}) \\
\end{array}
\right. \nn
\end{eqnarray}
Here $\tau \sim \alpha_s(m_b)$ is a radiative correction and $\delta_{\rm 
fact}$
denotes factorizable corrections which break the form factor relations. 
Precise measurements of the position of the zero $q_0^2$ can
give direct information about new physics effects through the
values of the Wilson coefficient $C_9^{\rm eff}$ (with $C_7^{\rm eff}$ 
determined from $B\to X_s\gamma$ decays).

The branching ratios of the $B\to K^* \ell^+\ell^-$ exclusive modes have 
been measured \cite{expexclusive} with the results
\begin{eqnarray}
&& {\cal B}(B\to K^* \ell^+ \ell^-) = \\
&& = \left\{
\begin{array}{cc}
(7.8^{+1.9}_{-1.7}\pm 1.2) \times 10^{-7} & \mbox{(BABAR)} \nonumber \\
(16.5\pm 2.3 \pm 0.9 \pm 0.4) \times 10^{-7} &\mbox{(BELLE)}\\
\end{array}\right.\nn
\end{eqnarray}
Differential distributions of the $q^2$-spectrum are also available,
as binned branching ratios.  First measurements of the forward-backward
asymmetry  $A_{FB}(q^2)$ have
been presented by the BELLE Collaboration \cite{expexclusive}, but due to 
large errors the position and even existence of a zero are still inconclusive.

\OMIT{
\begin{figure}[b!]
\begin{tabular}{cc}
{\includegraphics[height=5cm]{afbkstarll.eps}}
\end{tabular}
\caption{\label{fig1} 
Experimental data on the forward-backward asymmetry in $B\to K^*\ell^+\ell^-$
decays from BELLE~\cite{expexclusive}.
}
\end{figure}
}

In practice, the $K^*$ is always observed through its strong decay products 
$K^* \to K\pi$. We point out that this has several interesting 
implications. The multibody factorization relation proven here contains
a new factorizable contribution 
to the decay amplitude of leading order in $\Lambda/m_b$, which is not
present in the $B\to K^* \ell^+\ell^-$ factorization relation.
This introduces a shift in the position of the zero of the FBA in this region. 
As an application of the new factorization
relations, we compute the correction to the position of the zero arising
from this effect.

Another novel effect is the existence of a zero of the FBA also for a nonresonant 
$K\pi$ pair, which occurs at a certain dilepton invariant mass 
$q_0^2(M_{K\pi})$ depending on the hadronic invariant mass $M_{K\pi}$. 
In principle, this extends the applicability of the
SM test using the zero of the FB asymmetry also to nonresonant 
$B\to K\pi \ell^+\ell^-$ decays. In practice however, the calculation
of the position of the zero is complicated by the appearance of additional
nonperturbative contributions to the amplitude. We estimate the unknown 
nonfactorizable amplitude in $B\to K\pi \ell^+\ell^-$ in terms of a $K^*$ resonant
model.

We propose an alternative test of the SM using the {\em slope}  of the
$q_0^2(M_{K\pi}^2)$ curve, which can be shown to contain the same information about
the Wilson coefficients as the location of the zero itself. In contrast to the
absolute position of the zero, which depends on less well known hadronic parameters,
the slope of  the zero curve can be shown to be less sensitive to such effects.

In Sec.~II we introduce the SCET formalism and write down the effective 
Lagrangian for the rare $B\to X_s e^+e^-$ decay. Sec.~III presents the factorization
relations for the $B\to K \pi \ell^+\ell^-$ helicity amplitudes in the kinematical
region with a soft pion and a hard kaon. 
Sec.~IV lists the expressions for distributions in these decays, and gives
a qualitative discussion of the zero of the FBA in the nonresonant region.
Sec.~V contains a numerical analysis of the asymmetry, and finally Sec.~VI
summarizes our results.

\section{SCET formalism}
 
In the Standard Model the $\Delta S=1$ rare $B\to X_s \ell^+\ell^-$ decays are mediated 
by the weak Hamiltonian
\begin{eqnarray}\label{Heff}
H_W = -\frac{G_F}{\sqrt2} \lambda_t^{(s)} \sum_{i=1}^{10} C_i O_i(\mu)
\end{eqnarray}
with $\lambda_q^{(s)} = V_{tb} V^*_{ts}$.
The dominant contributions come from the radiative penguin 
$O_{7} = \frac{em_b}{4\pi^2 }\bar s\sigma_{\mu\nu} P_R F^{\mu\nu} b$
and the two operators containing the lepton fields $\ell = e, \mu$
\begin{eqnarray}
O_9 = \frac{\alpha}{\pi}(\bar s \gamma_\mu P_L b)(\bar \ell \gamma^\mu \ell)\,,\quad
O_{10} = \frac{\alpha}{\pi}(\bar s \gamma_\mu P_L b)(\bar \ell \gamma^\mu\gamma_5 \ell) 
\,.
\end{eqnarray}
We use everywhere in this paper the operator basis for $O_{1-6}$ defined in Ref.~\cite{CMM}.
Smaller contributions to the amplitude arise from T-products of the operators in
Eq.~(\ref{Heff}) with the electromagnetic current.

We choose the kinematics of the decay such that the total dilepton
momentum $q_\mu = (p_{\ell^+} + p_{\ell^-})_\mu$ points along the 
$-\vec e_3$ direction, and has the components $q = (q^0, 0, 0, -|\vec q\,|)$, expressed
in usual four-dimensional coordinates $a^\mu = (a^0, \vec a)$.
The hadronic system moves in the 
opposite direction $+\vec e_3$ in the $B$ rest frame. We define the light-cone unit 
vectors $n^\mu = (1,0,0,1), \bar n^\mu =(1,0,0,-1)$.
They can be used to project any vector $a^\mu$ onto light-cone directions, according to
$a_+ = n_\mu a^\mu$ and $a_- = \bn_\mu a^\mu$. Finally, we introduce a
basis of orthogonal unit vectors $\varepsilon_\pm
= \frac{1}{\sqrt2}(0,1,\mp i, 0), \varepsilon_0 = 1/\sqrt{q^2}(|\vec q|,0,0,q_0)$.

We will be interested in the kinematical
region with $q^2 \ll m_b^2$, for which the hadronic 
system has a large light-cone momentum component along $n$.
This defines the hard scale  $Q \equiv \bar n\cdot p_X  \sim m_b 
\gg \Lambda$, with $\Lambda \sim 500 $ MeV the typical scale
of the strong interactions.

The effective Hamiltonian Eq.~(\ref{Heff}) is matched in the SCET$_{I}$ onto
\begin{eqnarray}\label{HeffSCET}
H_W = -\frac{G_F}{\sqrt2} \lambda_t^{(s)} \frac{\alpha}{\pi}
\left\{ (\bar \ell\gamma_\mu \ell) J_V^\mu + (\bar \ell\gamma_\mu\gamma_5 \ell) J_A^\mu \right\}
\end{eqnarray}
where the currents $J_{V,A}^\mu$ have each the general form
\begin{eqnarray}\nn
 J^\mu_i &=& 
  c_1^{(i)}(\omega)\, \bar q_{n,\omega} \gamma_\perp^\mu P_L\, b_v\\
&& +\,
[c_2^{(i)}(\omega) v^\mu +  c_3^{(i)}(\omega) n^\mu ]\,\, \bar q_{n,\omega} P_R\, b_v \\
\label{JeffSCET}
&+&  b_{1L}^{(i)}(\omega_j)\, J^{(1L)\mu}(\omega_j) 
  + b_{1R}^{(i)}(\omega_j)\, J^{(1R)\mu}(\omega_j) \nn \\
&& + \, [b_{1v}^{(i)}(\omega_j) v^\mu + b_{1n}^{(i)}(\omega_j) n^\mu ]\,\, J^{(10)}(\omega_j)  \nn
\end{eqnarray}
with $i=V,A$, and integration over $\omega_j = (\omega_1, \omega_2)$ is implicit on the right-hand side.
This expansion contains the most general operators up to order $O(\lambda)$,
with $\lambda^2 = \Lambda/Q$.
The subleading operators are defined as
\begin{eqnarray}
J^{(1L,1R)}_\mu(\omega_1,\omega_2) &=& \bar q_{n,\omega_1}\, 
\Gamma^{(1L,1R)}_{\mu\alpha}
  \Big[\frac1{\bar n\cdot {\cal P}} ig {\cal B}^\alpha_{\perp n}\Big]_{\omega_2}
  b_v , \nn\\
J^{(10)}(\omega_1,\omega_2) &=& \bar q_{n,\omega_1}\, 
  \Big[\frac1{\bar n\cdot {\cal P}} ig \Bslash^\perp_n\Big]_{\omega_2}
  P_L\, b_v , 
\end{eqnarray}
with $\{\Gamma^{(1L)}_{\mu\alpha}\,, \Gamma^{(1R)}_{\mu\alpha}\} =
\{\gamma^\perp_\mu \gamma^\perp_\alpha P_R\,, \gamma^\perp_\alpha \gamma^\perp_\mu P_R\}$.
The collinear gauge invariant fields are defined as $q_n = W^\dagger \xi_n$, $ig {\cal B}_\mu = W^\dagger
[\bar n\cdot iD_c, iD_{c\mu}^\perp] W$, with $\xi_n$ the collinear quark field and 
$W = \exp[-g (\bn\cdot A_{n,q})/(\bn\cdot q)]$ a Wilson line of the collinear
gluon field.
We use throughout the notations of Ref.~\cite{ps1} with
$n\cdot v = \bn\cdot v = 1$.

The Wilson coefficients of the leading order SCET operators appearing in the 
matching of 
$J_{V,A}^\mu$ are \cite{scet}
\begin{eqnarray}
&& c_1^{(V)}(\omega,\mu) = \left(
C_9^{\rm eff} + 2m_b(\mu) \frac{n\!\cdot\! q}{q^2} C_7^{\rm eff}(\mu) (1+\tau(\omega, \mu)) \right)\nn \\ 
\label{c1V}
&&\qquad\times (1 - \frac{\alpha_s C_F}{4\pi} f_v(\omega,\mu)) 
+ O(\alpha_s^2(Q)) \\
\label{c1A}
&& c_1^{(A)}(\omega,\mu) = C_{10}(1 - \frac{\alpha_s C_F}{4\pi} f_v(\omega,\mu)) + O(\alpha_s^2(Q))
\end{eqnarray}
The effective Wilson coefficients $C_{7,9}^{\rm eff}$ include the contributions of the
operators $O_{1-6}$ and $O_8$. For convenience they are listed in the Appendix, together with the
functions $f_v(\omega,\mu)$ containing the $O(\alpha_s(Q))$
contribution to the Wilson coefficient of the vector current in SCET, and
$\tau(\omega,\mu)$ giving the additional contribution from the tensor current.

The Wilson coefficients of the $O(\lambda)$ SCET$_{\rm I}$ operators are given at
leading order in $\alpha_s(Q)$ by
\begin{eqnarray}\label{b1LO}
&& b^{(V)}_{1L}(\omega_1,\omega_2) = - \frac{2n\cdot q}{q^2} \Big[C_7^{\rm eff}  
+ \frac{\bar x \omega}{8m_b} e_u \bar C_2 t_\perp(x, m_c) 
\Big] \nn \\
&& b^{(V)}_{1R}(\omega_1,\omega_2) = \Big( 2m_b(\mu) \frac{\bn\cdot q}{q^2} C_7^{\rm eff} + 
C_9^{\rm eff} \Big) \frac{1}{\omega} \\
&& \qquad + \frac{2n\cdot q}{q^2} \Big(
\frac{\bar x\omega}{8m_b} e_u \bar C_2 t_\perp(x, m_c)  \Big) \nn \\
&& b^{(A)}_{1L}(\omega_1,\omega_2) = 0 \\
&& b^{(A)}_{1R}(\omega_1,\omega_2) = C_{10}\frac{1}{\omega}
\end{eqnarray}
with $\omega_1 = x \omega\,, \omega_2 = - \bar x \omega$, and $\bar x = 1-x$.
We neglect here smaller contributions from the operators $O_{3-6}$ and the 
gluon penguin $O_8$, which will be retained only in the leading order SCET 
Wilson coefficients $c_1^{(V,A)}(\omega,\mu)$. 
The complete expression can be extracted from Ref.~\cite{fbcor1}.
The Wilson coefficients $\bar C_i$
are defined in the Appendix. The function $t_{\perp}(x,m_c)$
appears in matching from graphs with both the photon and the transverse collinear
gluon emitted from the charm loop
\cite{fbcor1} and is given in Eq.~(\ref{tperp}) of the Appendix.

The coupling of the virtual photon $\gamma^*\to \ell^+\ell^-$ to the light 
quarks can also occur through diagrams with intermediate hard-collinear
quarks propagating along the photon momentum \cite{CK}. (Such a description
is appropriate only for a range of the dilepton invariant mass 
$q^2 \leq 1.5$ GeV$^2$ which can be considered hard-collinear, see below.)
Such contributions are mediated by new terms in the SCET$_{\rm I}$ effective 
Lagrangian, which in the SM contains only one operator at leading order
\begin{eqnarray}\label{spI}
{\cal H}_{\rm sp} &=& \frac{4G_F}{\sqrt2} \lambda_t^{(s)}\\
 & & \times \sum_{q=u,d,s}
b_{\rm sp}^{(q)}(\omega_j) (\bar q_{\bn,\omega_3} \nslash
P_L b_v )(\bar q_{n,\omega_1} \frac{\bnslash}{2} P_L s_{n,\omega_2} )\nn
\end{eqnarray}
The Wilson coefficient is given by
\begin{eqnarray}
b_{\rm sp}^{(q)}(z) &=& \frac{\lambda_u^{(s)}}{\lambda_t^{(s)}} 
(\bar C_2+\frac{\bar C_1}{N_c}) \delta_{qu} -
(\bar C_4+\frac{\bar C_3}{N_c})\\
& & + O(\alpha_s(Q))  \nn
\end{eqnarray}
where we neglect again smaller contributions proportional to $C_8$.

For application to exclusive $B\to M$ form factors, with $M=\pi\,, \rho\,,\cdots$ a light
meson, the SCET$_{\rm I}$ effective Lagrangian
in Eq.~(\ref{HeffSCET}) has to be matched onto SCET$_{\rm II}$ operators.
This requires taking into account the interaction with the soft spectator quark
in the B meson \cite{scet1, bpsff}. Working to leading order in SCET$_{\rm II}$, this matching
contains two types of operators~\cite{bpsff,ps1,befe,CK,hillff}
\begin{eqnarray}\label{scet12}
&& 
J_i^\mu \to 
c_1^{(i)}(\omega) O_{\rm nf}^\mu + [c_2^{(i)}(\omega) v^\mu +  c_3^{(i)}(\omega) n^\mu ]
O'_{\rm nf}  \nn \\
&& \hspace{2cm} + J^\mu_{\rm i,f} + \delta_{i,V} J^\mu_{\rm sp} + \cdots 
\end{eqnarray}
The first type are the so-called `nonfactorizable' operators, denoted here
as $O_{\rm nf}^\mu, O'_{\rm nf}$. They are defined such that they 
include the contributions of the leading operators
in the SCET$_{\rm I}$ Lagrangian.
The second type of operators are the so-called  
factorizable and spectator interaction 
operators, denoted as $J_{\rm f}^\mu$ and $J^\mu_{\rm sp}$, respectively.
Although their form is similar, they arise in matching from different
operators in SCET$_{\rm I}$, as follows. 
The operators $J_{\rm f}^\mu$ are obtained from 
the $O(\lambda)$ operators in the SCET$_{\rm I}$ current.
The spectator operators contribute only to $J_V^\mu$ in Eq.~(\ref{HeffSCET}),
and arise from the SCET$_{\rm I}$ weak nonleptonic
effective Hamiltonian Eq.~(\ref{spI}).
The ellipses denote terms suppressed by powers of $\Lambda/m_b$.

The matrix elements of the nonfactorizable operators in 
Eq.~(\ref{scet12}) corresponding to a $B\to M_n$ transition, 
are parameterized in terms of soft form factors. We
define them as \cite{befe}
\begin{eqnarray}\label{zetaBM}
&& \langle M_n(p_M)|O'_{\rm nf}|\bar B\rangle = 2E_M \zeta_0(E_M, \mu) \\
&&
\langle M_n(p_M)|O_{\rm nf}^\mu \varepsilon_{-\mu}^*|\bar B\rangle = 
2E_M \zeta_\perp(E_M)
\end{eqnarray}
where in the first matrix element  $M_n$ is a pseudoscalar meson,
and in the second $M_n$ is a transversely polarized vector meson.

The factorizable operators in Eq.~(\ref{scet12}) are nonlocal 
soft-collinear four-quark operators. As mentioned above, they
are of two types, denoted as factorizable-type (f), 
and spectator-type (sp). 
The $J^\mu_{\rm i,f}$ operators have the generic form 
\begin{eqnarray}\label{factgen}
&& J^\mu_{\rm i,f} \sim \int \mbox{d}x \mbox{d}z \mbox{d}k_+ 
b^{(i)}(z) J (x,z,k_+) \\
&& \hspace{1cm} \times (\bar q_{k+} \Gamma_S b_v)
(\bar q_{n,\omega_1} \Gamma_C q_{n,\omega_2}) \nn
\end{eqnarray}
where $b(z)$ are SCET$_{\rm I}$ Wilson coefficients, and
$J$ are jet functions. They are given in explicit form in Eq.~(\ref{JeffIIfact}) below.
We use a momentum space notation for the
nonlocal soft operator, defined by
\begin{eqnarray}
(\bar q_{k+}^i  b_v^j)(0) = \int \frac{d\lambda}{4\pi} 
e^{-\frac{i}{2} \lambda k_+} \bar q(\lambda\frac{n}{2}) Y(\lambda,0) b_v^j(0)
\end{eqnarray}
with $Y(\lambda,0)= P\exp(ig\int d\alpha n_\mu A^\mu (\alpha n/2))$ a 
soft Wilson  line along the direction $n_\mu$.

There are two jet functions, defined as Wilson coefficients appearing in the 
matching of T-products of the SCET$_{\rm I}$ currents Eq.~(\ref{JeffSCET}) with the 
ultrasoft-collinear subleading Lagrangian ${\cal L}_{q\xi}^{(1)}$,
onto SCET$_{\rm II}$ \cite{bpsff,ps1,bprs}
\begin{eqnarray}
&& T [\bar q_{n,\omega_1} ig {\cal B}_{n,\omega_2}^{\perp \alpha}]^{ia}(0)
[ig \Bslash_n^\perp q_n]_{\omega=0}^{jb}(y) =  \\
&& i\delta^{ab} \delta(y_+) \delta^{(2)}(y_\perp)
\frac{1}{\omega} \int_0^1 \mbox{d} x \int \frac{d k_+}{4\pi}
e^{ik_+ y_-/2} \nn\\
&& \times \{
J_\parallel(x,z,k_+) [(\nslash \gamma_\perp^\alpha)^{ji} [\bar q_{n,x\omega} 
\frac{\bnslash}{2} q_{n,-\bar x\omega}]\nn\\
&& \hspace{2cm} +
(\nslash\gamma_5 \gamma_\perp^\alpha)^{ji} [\bar q_{n,x\omega} \frac{\bnslash}{2}\gamma_5 q_{n,-\bar x\omega}] ] \nn\\
&& \quad + J_\perp(x,z,k_+) (\nslash \gamma_\perp^\alpha 
\gamma_\perp^\beta)^{ji}
[\bar q_{n,x\omega} \frac{\bnslash}{2} \gamma^\beta_\perp q_{n,-\bar x\omega}] \} \nonumber
\end{eqnarray}
with $\omega_1 = z\omega$ and
$\omega=\omega_1-\omega_2$. The jet functions 
are generated by physics at the hard-collinear scale $\mu_c^2 \sim Q\Lambda$,
and have perturbative expansions in $\alpha_s(\mu_c)$.
At lowest order in $\alpha_s(\mu_c)$ they are given by 
\cite{bpsff,ps1}
\begin{eqnarray}\label{jets}
J_{\perp,\parallel}(x,z,k_+) =
\frac{\pi \alpha_s C_F}{N_c} \delta(x-z) \frac{1}{\bar x k_+}
\end{eqnarray}

Finally, the spectator-type operators have the form 
\begin{eqnarray}\label{Heffspgen}
&& J^\mu_{\rm sp} = 
\int_0^1 \mbox{d}z b_{\rm sp}(z) \int dk_-  
J_{\rm sp}(k_- - \frac{q^2}{n\cdot q}) \\
&& \qquad \times (\bar q_{k_-} \gamma^\mu
\bnslash \nslash P_L b_v) (\bar s_{n,z\omega} \frac{\bnslash}{2} P_L q_{n,-\bar z\omega}) \nn
\end{eqnarray}
The jet function $J_{\rm sp}(k_-)$ is the same as the jet function
appearing in the factorization relation for $B \to \gamma\ell\bar\nu$.
It can be extracted from the results of Ref.~\cite{LPW}
and is given at one-loop order by
\begin{eqnarray} 
\label{Jsp-given}
J_{\rm sp}(k_-) = \frac{1}{k_-  + i\epsilon}\Big(1 + \frac{\alpha_s C_F}{4\pi}
(L^2 - 1 
- \frac{\pi^2}{6}) \Big)
\end{eqnarray}
with $L = \log[(-n\cdot q k_- - i\epsilon)/\mu^2]$.

The matrix elements of the factorizable and spectator operators in the
$\bar B\to M_n$ transition at large recoil are computed as convolutions of 
the product of collinear and soft matrix elements. 
Adding also the nonfactorizable contribution, the generic
form of the factorization relation for the hadronic matrix element for
$\bar B\to M_n$  + leptons is written as \cite{bpsff} (with $i=V,A$)
\begin{eqnarray}\label{Mfactgen}
&& \langle M_n |J_i |\bar B\rangle = c_j^{(i)} 2 E_M \zeta^{BM} \\
&& + \int \mbox{d}x \mbox{d}z \mbox{d}k_+ 
b^{(i)}(z) J_j (x,z,k_+) \nn \\
&& \hspace{1cm} \times \langle 0 |\bar q_{k+} \Gamma_S b_v |\bar B(v)\rangle
\langle M_n |\bar q_{n,\omega_1} \Gamma_C q_{n,\omega_2} |0\rangle \nn\\
&& + \int \mbox{d}x b_{\rm sp}(x) \phi_M(x) \int  \mbox{d}k_-
J_{\rm sp} (k_-) \langle 0 | \bar q_{k_-} \Gamma_S b_v|\bar B\rangle \nn\,.
\end{eqnarray}

The nonperturbative soft and collinear matrix elements appearing in this relation
are given by the B-meson and light meson light-cone wave functions, respectively. 
We list here their expressions, adopting the following
phase conventions for the meson states
\begin{eqnarray}
&& (\pi^+, \pi^0, \pi^-) = (u\bar d, \frac{1}{\sqrt2}(u\bar u - d\bar d) , d\bar u ) \\
&& (B^-, \bar B^0) = (b\bar u, b\bar d) \,. \nn
\end{eqnarray}
The light mesons' light cone wave functions are given by
\begin{eqnarray}
&& \langle \bar K_n(p_K) | \bar s_{n,\omega_1} \bnslash P_L q_{n,\omega_2}| 0\rangle =
\frac{i}{2} f_K \bn\!\cdot\! p_K \phi_K(x) \\
&& \langle \bar K^*_n(p_{K^*},\eta) | \bar s_{n,\omega_1} 
\bnslash \gamma_\perp^\mu q_{n,\omega_2}| 0\rangle =
\frac12 f_{K^*}^T \bn\!\cdot\! p_{K^*} \eta^{*\mu}_\perp \phi_{K^*}^\perp(x) \,.\nn
\end{eqnarray}
and
the B meson light-cone wave function is defined as \cite{befe}
\begin{eqnarray}\label{Bwf}
&& \langle 0 | \bar u^i_{k_+} b_v^j|B^-(v)\rangle = \\
&& -\frac{i}{4}f_B m_B \left\{
\frac{1+\vslash}{2} [\bnslash  \phi_+^B(k_+) +
\nslash \phi_-^B(k_+) ]\gamma_5 \right\}_{ji} \nonumber
\end{eqnarray}
The matrix element of the $\bar q_{k_-} b_v$ appearing in the last term of
Eq.~(\ref{Mfactgen}) 
can be obtained from this by the substitution $n_\mu \leftrightarrow \bn_\mu$.

\subsection{Factorization in multibody B decays}

\begin{figure}
{\includegraphics[height=1.8cm]{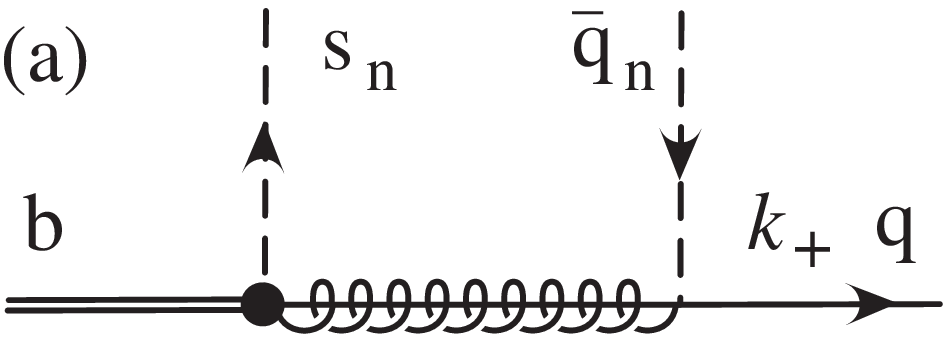}}
{\includegraphics[height=1.8cm]{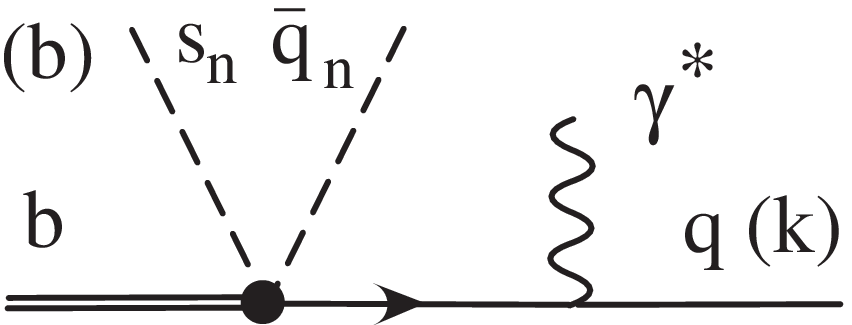}}
\caption{\label{fig4} 
Leading order SCET$_{\rm I}$ graphs contributing to 
the factorizable amplitude for $\bar B \to \bar K_n \pi $ + leptons. 
a) the T-product $\{J_{V,A}^\mu\,, i{\cal L}_{q\xi_n}^{(1)}\}$, 
where the filled circle represents the SCET current $J_{V,A}^\mu$; b)
spectator-type contribution, with the virtual photon attaching to 
the light current quark; the intermediate quark propagator can be
either hard-collinear along the photon direction $\bn_\mu$, or 
hard, depending on the ratio $n\cdot q \Lambda/q^2$ being larger or 
less than 1, respectively.
After matching onto SCET$_{\rm II}$ these graphs contribute
to the operators $J_{\rm i,f}$ and $J_{\rm sp}$, respectively.
}
\end{figure}

We consider here the application of the SCET formalism to
rare $\bar B \to \bar K_n \pi \ell^+\ell^-$ decays into final states 
containing one energetic hadron $\bar K_n$ and a soft hadron $\pi$.
The heavy-light currents in Eq.~(\ref{JeffSCET}) contribute to such
processes again through T-products 
with the ultrasoft-collinear subleading Lagrangian \cite{scet1}. Typical
diagrams in SCET$_{\rm I}$ contributing to these T-products are shown in 
Fig.~\ref{fig4}.

After integrating out the modes with virtuality $p_{hc}^2 \sim \Lambda Q$ 
connected with the hard-collinear degrees of freedom, these T-products
are matched onto SCET$_{\rm II}$ \cite{bpsff}.
Performing a Fierz transformation of the four-quark operators, one finds
the following result for the factorizable operators 
\begin{eqnarray}\label{JeffIIfact}
&& J^\mu_{\rm i, fact} =  -\frac{1}{2\omega} 
\int \mbox{d}x \mbox{d}z \mbox{d}k_+ b_{1L}^{(i)}(z) J_\perp(x,z,k_+) \\
&&\qquad\qquad \times 
(\bar q_{k_+} \nslash \gamma^\perp_\mu \gamma_\perp^\lambda P_R  b_v)
(\bar s_{n,\omega_1} 
\frac{\bnslash}{2} \gamma_\perp^\lambda q_{n,\omega_2}) \nn\\
&&
- \frac{1}{2\omega}
\int \mbox{d}x \mbox{d}z \mbox{d}k_+ b_{1R}^{(i)}(z) J_\parallel(x,z,k_+) \nn \\
&&\qquad\qquad \times 
(\bar q_{k_+} \nslash \gamma^\perp_\mu P_R  b_v)(\bar s_{n,\omega_1} 
\frac{\bnslash}{2} P_L q_{n,\omega_2}) \nn \\
&&
- \frac{1}{\omega}
\int \mbox{d}x \mbox{d}z \mbox{d}k_+ [b_{1v}^{(i)}(z) v_\mu + b_{1n}^{(i)}(z) n_\mu)]\nn\\
&& \qquad \times J_\parallel(x,z,k_+)
 (\bar q_{k_+} \nslash  P_L b_v)(\bar s_{n,\omega_1} 
\frac{\bnslash}{2} P_L q_{n,\omega_2}) \nn 
\end{eqnarray}
The coefficients $b_j(z)$ are related to the
Wilson coefficients $b_j(\omega_1,\omega_2)$ of Eqs.~(\ref{b1LO})
as $b_j(z) \equiv b_j((1-z)\omega, z\omega)$;
the labels of the collinear fields are parameterized as 
$\omega_1 = x\omega, \omega_2 = -\omega (1-x), 
\omega = \omega_1 - \omega_2 = \bn\cdot p_M$, with $p_M$ the
momentum of the collinear meson $M_n$ produced by the
collinear part of the operator.

Finally, the spectator-type factorizable operators arise from
diagrams where the photon attaches to the light current quark 
Fig.~\ref{fig4}b.  Although the spectator quark is not 
involved in these contributions, we will continue to use the
same terminology as in the $B\to M_n$ case, due to the similarity of 
the corresponding operators.

The effective theory treatment of these contributions depends on the 
relative size of the virtuality of the photon $q^2$ and the typical 
hard-collinear scale $m_b \Lambda$. These two scales correspond to the
two terms $k_- + q^2/n\cdot q + i\varepsilon$ in the propagator of the
intermediate quark in Fig.~\ref{fig4}b. 

Several approaches are used in the literature to deal with these 
contributions, which we briefly review in the following. 
One possible approach, used in \cite{fbcor1,fbcor3} in QCD factorization, 
is to keep both terms in the propagator, and not expand in their ratio.
From the point of view of the effective theory, this approach 
is equivalent to treating the photon as a hard-collinear mode
moving along the photon $\bn_\mu$ direction \cite{CK}. This approach is 
certainly appropriate for real photons, and for hard-collinear photons
$q^2 \sim 1.5$  GeV$^2$. It is not clear whether it can be also applied to 
photons with $q^2 \sim 4 $ GeV$^2$, as is the case here. 

In this approach the spectator-type effective Hamiltonian Eq.~(\ref{spI}) 
contributes to exclusive decays through T-ordered products with the leading
order SCET$_{\rm I}$ Lagrangian describing photon-quark couplings 
\cite{LPW,CK}.
After matching onto SCET$_{\rm II}$, these T-products are matched onto
one single operator, which can be written as an addition to $J_V^\mu$,
and is given by
\begin{eqnarray}
&& J^\mu_{\rm sp} =  \frac{8\pi^2}{q^2} \sum_{q=u,d,s} e_q 
\int_0^1 \mbox{d}z b_{\rm sp}^{(q)}(z) \int dk_-  J_{\rm sp}(k_- - \frac{q^2}{n\cdot q}) \nn\\
\label{Heffsp}
&& \qquad \times (\bar q_{k_-} \gamma_\mu
\bnslash \nslash P_L b_v) (\bar s_{n,\omega_1} \frac{\bnslash}{2} P_L q_{n,\omega_2}) \
\end{eqnarray}
For consistency with the other factorizable operators included, we work to
tree level in $\alpha_s(Q)$, but keep terms of $O(\alpha_s(\mu_c))$ in the 
matrix elements of the factorizable operators. 

In the kinematical region we are interested in, a more appropriate treatment
of these contributions makes use of an expansion in powers of
$n\cdot q \Lambda/q^2 \sim 0.37 $. This is similar to the approach adopted in 
Ref.~\cite{bdi} for weak annihilation contributions to $B\to \pi\ell^+\ell^-$. 
The SCET$_{\rm II}$ operators obtained in this way are similar to those in
Eq.~(\ref{Heffsp}), except that the soft operator is local. Keeping terms to 
second order in $n\cdot q \Lambda/q^2$, the spectator operator in this approach
reads
\begin{eqnarray}\label{Heffsplocal} 
&& J^\mu_{\rm sp} =  \frac{16\pi^2}{q^2} \sum_q e_q
\Big\{ \frac{q^\alpha}{q^2} (\bar q \gamma^\mu \gamma_\alpha \nslash  P_L b)\\
&& \qquad +
\Big( \frac{g^{\alpha\beta}}{q^2} - \frac{2q^\alpha q^\beta}{q^4} \Big)
(\bar q \gamma^\mu \gamma_\alpha (-i \stackrel{\leftarrow}{D}_\beta) \nslash  
P_L b) \Big\} \nn \\
&& \qquad \times \int_0^1 \mbox{d}z b_{\rm sp}^{(q)}(z) 
(\bar s_{n,\omega_1} \frac{\bnslash}{2} P_L q_{n,\omega_2}) \nn
\end{eqnarray}

In this paper we will adopt the latter
approach to the treatment of the spectator amplitude, working at leading 
order in $n\cdot q \Lambda/q^2$. We quote our results in terms of the first
approach, which has an additional convolution over $k_-$.
However, it is straightforward to translate between the two approaches, 
simply by replacing
\begin{equation}\label{subst}
J_{\text{sp}}(k_-)  \leftrightarrow 
\frac{n\cdot q}{q^2}
\end{equation}
below (in, e.g., Eqs.~(\ref{Hsp}) and (\ref{hspdef})). The advantage
of this approach is that the predictions are independent on the
details of the matrix elements of the nonlocal soft operator (the B meson 
light-cone wave functions), but can be computed exactly in the soft pion
limit to second order in the $n\cdot q \Lambda/q^2$ expansion.

\begin{figure}
\centering
\begin{picture}(300,170)
\put(10,20){\includegraphics[height=4.5cm]{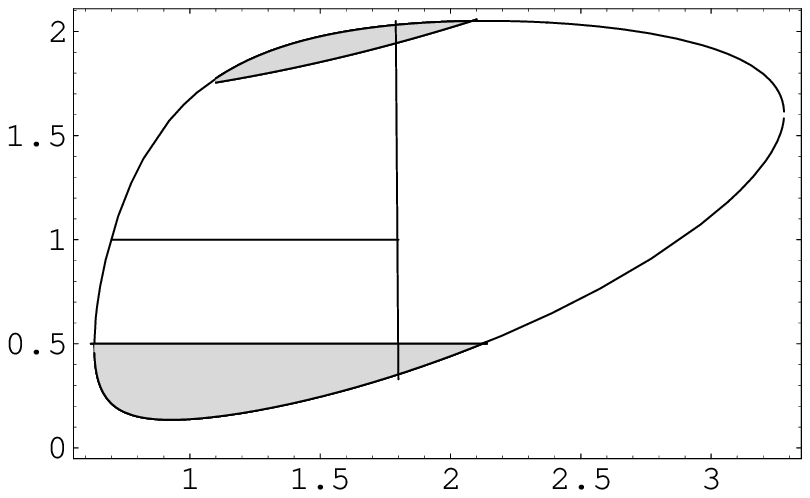}}
\put(-5,145){
{\large
$E_\pi$
}}
\put(-15,125){
{\large
(GeV)
}}
\put(80,-5){
{\large
$M_{K\pi}\mbox{ (GeV)}$
}}
\put(72,65){
{\large
$\mbox{ I}$
}}
\put(74,95){
{\large
$\mbox{ II}$
}}
\put(40,130){
{\large
$\mbox{ III}$
}}
\end{picture}
\caption{\label{figdal} 
The phase space of the decay $B\to K\pi \ell^+ \ell^-$ at $q^2 = 4$ GeV$^2$,
in variables $(M_{K\pi}, E_\pi)$.
The 3 regions shown correspond to: (I) soft pion $E_\pi \sim \Lambda$;
the shaded region $E_\pi\leq 0.5$ GeV
shows the region of applicability of chiral perturbation theory; (II)
collinear pion and kaon $E_\pi \sim Q,
E_K > 1$ GeV; 
(III) soft kaon $E_K < 1$ GeV.
}
\end{figure}

In order to have a clean power counting
of the transition amplitudes, we divide the phase space of the
$B\to K\pi\ell^+\ell^-$ decay into several regions,
shown in Fig.~\ref{figdal}:

I) the region with one soft pion and one energetic
kaon $B\to K_n \pi_S$, $E_\pi \sim \Lambda, E_K \sim Q$ and
$M_{K\pi}^2 \sim \Lambda Q$. This region will be the main
interest of our paper.

II)
the region $B\to (K_n \pi_n)_{K^*}$ describing decays into
an energetic $K\pi$ pair with a small invariant mass $M_{K\pi}\sim
\Lambda$.  This is dominated by one-body decays into a collinear
meson $B\to K^*_n$, followed by $K_n^* \to K_n\pi_n$. This region
will be treated essentially the same way as a one-body decay.

III) the region with a soft kaon and an energetic pion $E_K \sim \Lambda, E_\pi
\sim Q$. The decay amplitude in this region is suppressed by $\Lambda/Q$
relative to that in the other two regions I, II, and will be neglected in the
rest of the paper.

We will use the SCET formalism described above to derive a factorization 
relation in the region (I).
The matrix elements of the nonfactorizable operators are
parameterized in terms of soft nonperturbative matrix elements,
in analogy with the $B\to M_n$ transition.
We define them as complex functions of the momenta of the final
state hadrons, with mass dimension zero 
\begin{eqnarray}\label{zeta}
&&
\langle M_n M'_S|O_{\rm nf}^\mu \varepsilon_{-\mu}^*|\bar B\rangle = 
\zeta_\perp^{BMM'}(E_M,p_{M'})\\
&& \langle M_n M'_S|O'_{\rm nf}|\bar B\rangle = \zeta_0^{BMM'}(E_M,p_{M'})
\nn
\end{eqnarray}

The matrix elements of the factorizable and spectator-type
operators are given again by convolutions as in Eq.~(\ref{Mfactgen}),
with a different soft matrix element
\begin{eqnarray}\label{Mfactgen1}
&& \langle M_n M'_S |J_{\rm i} |\bar B(v)\rangle \sim 
\int \mbox{d}x \mbox{d}z \mbox{d}k_+ 
b^{(i)}(z) J_j (x,z,k_+) \nn \\
&& \hspace{0.5cm} \times \langle M'_S |\bar q_{k+} \Gamma_S b_v |\bar B(v)\rangle
\langle M_n |\bar q_{n,\omega_1} \Gamma_C q_{n,\omega_2} |0\rangle \\
&& + \int \mbox{d}x b_{\rm sp}(x) \phi_M(x) \int  \mbox{d}k_-
J_{\rm sp} (k_-) \langle 0 | \bar q_{k_-} \Gamma_S b_v|\bar B\rangle \nn\,.
\end{eqnarray}

These factorization relations contain several new hadronic nonperturbative
matrix elements, which we define next.
The new $\bar B\to \pi$ soft matrix element 
is defined in terms of the soft operator
\begin{eqnarray}
O_{\mu}(k_+) = \int \frac{d\lambda}{4\pi} 
e^{-\frac{i}{2}\lambda k_+}  \bar u(\lambda \frac{n}{2}) Y_n(\lambda,0) 
\gamma^\perp_\mu \frac{\nslash}{2} P_R b_v(0)
\end{eqnarray}
appearing in the term in Eq.~(\ref{JeffIIfact}) proportional to $b_{1R}(z)$,
and in the spectator operator. In the latter, one has to take into account that 
the light-like separation between the fields is along the direction $\bn_\mu$,
rather than $n_\mu$ as in the factorizable operators $J^\mu_{\rm i, fact}$.

The matrix element of the operator $O_\mu$ defines a soft function $S$ as
\begin{eqnarray}\label{Sdef}
\langle \pi^+\!(p_\pi) | O_{\mu}\! (k_+) |\bar B^0\!(v) \rangle \!=\!
-(g^\perp_{\mu\nu} - i \varepsilon^\perp_{\mu\nu}) p^\nu_\pi
S(k_+, t^2, p_\pi^+)\nn\\
\end{eqnarray}
with $t = m_B v - p_\pi$.
For simplicity of notation, we will drop the kinematical arguments 
of the soft function $S(k_+, t^2, \zeta)$ whenever no risk of confusion
is possible, and show explicitly only its 
dependence on the integration variable $k_+$.
The matrix elements of the spectator operator in Eq.~(\ref{Heffsp})
are obtained from Eq.~(\ref{Sdef}), with the replacements $n \leftrightarrow \bn$ and 
$\varepsilon^\perp_{\mu\nu} \to
- \varepsilon^\perp_{\mu\nu}$.

The function $S(k_+)$ is the B physics
analog of a generalized parton distribution function (GPD), commonly
encountered in nucleon physics \cite{GPD}.
The support of this function is the range $-n\cdot p_\pi \leq
k_+ \leq \infty$, and its physical interpretation is different
for positive and negative values of $k_+$. For $k_+ > 0$ 
(the resonance region) the soft function gives the amplitude of
finding a $u\bar d$ pair in the $\bar B^0$ meson, while for
$k_+ < 0$ (the transition region), the soft function gives the
amplitude for the $b\to u$ transition of the $\bar B^0$ meson
into a $\pi^+$ meson. The soft function $S(k_+)$ is continuous at
the transition point $k_+=0$ \cite{GPD}, which is important for ensuring the
convergence of the $k_+$ convolutions in the factorization relation
Eq.~(\ref{factmulti}).

We recall here the main properties of the soft function $S(k_+)$,
which were discussed in Ref.~\cite{FK,rad}.
Time invariance of the strong interactions constrains it to be real. 
Its zeroth moment with respect to $k_+$ is given by 
\begin{eqnarray}\label{Snorm}
\int_{-p_\pi^+}^\infty dk_+ S(k_+, t^2, p_\pi^+) = -\frac14 n\cdot p f_T(t^2)
\end{eqnarray}
with $f_T(t^2)$ the $B\to \pi$ tensor form factor 
defined as 
\begin{eqnarray}\label{fTdef}
\langle \pi(p') |\bar q i\sigma_{\mu\nu} b | \bar B(p)\rangle =
f_T(t^2) (p_\mu p'_\nu - p_\nu p'_\mu)
\end{eqnarray}
Its $N$-th moments with
respect to $k_+$ are related in a similar way to $\bar B\to \pi$ form factors of
dimension $3+N$ heavy-light currents of the form $\bar q (n\cdot iD)^{N} b_v$ \cite{rad}.

In the soft pion region, chiral symmetry can be used to
relate $S(k_+, t^2, \zeta)$ in the region $k_+ > 0$ to one of the B 
meson light-cone wave functions $\phi_+^B(k_+)$ defined in Eq.~(\ref{Bwf}),
according to \cite{GPchiral}
\begin{eqnarray}\label{SChPT}
S(k_+, t^2, \zeta) = \frac{g f_B m_B}{4f_\pi}
\frac{1}{v\cdot p_\pi + \Delta} \phi_+^B(k_+)
\end{eqnarray}
Here $g$ is the $BB^*\pi$ coupling appearing in the leading order heavy hadron
chiral effective Lagrangian \cite{wise,BuDo,TM,review}.
No such constraint is obtained using chiral symmetry for $S(k_+,t^2,\zeta)$ in 
the transition region ( $k_+ <0$).

Collecting all the contributions, the amplitude for $B\to M_n M'_S$ + leptons is given by
a sum of
factorizable and nonfactorizable terms, corresponding to the matrix
elements of the SCET$_{\rm II}$ operators in Eq.~(\ref{scet12}). This leads
to a factorization relation for such processes, which can be
written schematically as 
\begin{eqnarray}\label{factmulti}
&& A(B\to [M_n M'_S]\mbox{+leptons} ) = c_i(\bar n \cdot p_M) \zeta^{BMM'} \\
&& \quad +
\int dz \int dx dk_+ b_i(z) J_j(x, z, k_+) \phi_M(x) S(k_+; p_{M'}) \nonumber \\
&& \quad +
\int dx\, b_{\rm sp}(x) \phi_M(x) \int dk_+ J_{\rm sp}(k_+) S(k_+; p_{M'}) \nonumber 
\end{eqnarray}
This factorization relation 
has several important properties \cite{GPchiral}.
First, the nonfactorizable contributions to the decay amplitudes of semileptonic
and radiative decays satisfy symmetry relations following from the universality 
of the soft matrix element $\zeta_\perp^{BK\pi}$. They contribute only to the
decays $\bar B\to [M_n M'_S]_{h=-1}\ell^+\ell^-$ into final hadronic
states with total helicity $-1$. Second, the amplitude for $+1$ helicity is factorizable, and given by
a convolution as seen in the second term of Eq.~(\ref{factmulti}). 
Finally, the factorizable terms contain a new source of strong phases,
arising from the region $k_+\leq 0$ where the jet function develops a nonzero
absorbtive part. This represents a new, factorizable, mechanism for generating final state
rescattering.

Treating the spectator amplitudes in an expansion in powers of $n\cdot q\Lambda/q^2$
according to Eq.~(\ref{Heffsplocal}), the soft matrix elements are given by
$B\to \pi$ form factors of dimension-3 and 4 local operators. The leading order term
contains the form factors of the vector current
\begin{eqnarray}
\langle \pi(p_\pi)|\bar u \gamma_\mu b|\bar B(p)\rangle \!\!=\!\!
f_+(t^2) (p+p_\pi)_\mu\!\! +\!\! f_-(t^2) (p\! -\! p_\pi)_\mu
\end{eqnarray}
The $f_\pm(t^2)$ form factors appear in the matrix element of Eq.~(\ref{Heffsplocal})
in the combination $f_+ - f_-$. 
In the hard photon approach, the last term of Eq.~(\ref{factmulti})
has the form $A_{\rm sp} \sim f_T(t^2) \int_0^1 dx\, b_{\rm sp}(x) \phi_M(x)$, which follows
from making the substitution Eq.~(\ref{subst}) in this relation, and using Eq.~(\ref{Snorm}).

At subleading order in $n\cdot q\Lambda/q^2$, the form factors of dimension-4
currents $\bar u \gamma_\alpha iD_\beta  b$ are also needed. They can be computed in the
soft pion limit using chiral perturbation theory methods as discussed in Ref.~\cite{GP1}.

In the following section
we derive the detailed form of these factorization relations for the
$\bar B\to K\pi\ell^+\ell^-$ decays.

\section{Factorization relations for $\bar B\to K\pi \ell^+\ell^-$}
  
The decay amplitudes $\bar B\to \bar K_n \pi\ell^+\ell^-$ into an energetic
kaon and one soft pion can be parameterized in terms of 6 independent
helicity amplitudes $H^{(V,A)}_\lambda(\bar B\to \bar K_n \pi)$ with $\lambda
= \pm 1, 0$. They are defined as the matrix elements of the two
hadronic currents in Eq.~(\ref{HeffSCET})
\begin{eqnarray}\label{Hdef}
H^{(V,A)}_\lambda(\bar B\to \bar K_n \pi) = 
\varepsilon_\lambda^{\mu *}\langle \bar K \pi|
J_{V,A}^\mu |\bar B(v)\rangle
\end{eqnarray}

Working at leading order in $1/m_b$, the helicity amplitudes
can be written as a sum of nonfactorizable and factorizable terms,
arising from the corresponding SCET$_{\rm II}$ operators in Eq.~(\ref{scet12})
\begin{eqnarray}\label{Hsum}
H^{(V,A)}_\lambda(B\to K_n \pi) \!=\!\!\!\! \sum_{i=\rm nf,f, sp}\!\!
H^{(V,A),i}_\lambda(B\to K_n \pi) 
\end{eqnarray}
The three contributions to each helicity amplitudes are computed as described
in Sec.~II. The nonfactorizable terms are given in terms of the soft functions
$\zeta_i^{BK\pi}$ defined in Eq.~(\ref{zeta}), and the factorizable and 
spectator contributions
are given by factorization relations of the form shown in Eq.~(\ref{factmulti}).
In this section we present explicit results for the transverse
helicity amplitudes.

We start by recalling the results for the one-body decays $B\to K^*_n \ell^+\ell^-$. The 
factorization relations for this case are well-known
\cite{fbcor1,bpsff,ps1,Beneke:2003pa,CK,pol} and are given by
(with $i=V,A$)
\begin{eqnarray}\label{H+I}
&& H_+^{(i)}(\bar B\to \bar K^*_n) = 0\\
\label{H-I}
&& H_-^{(i)}(\bar B\to \bar K^*_n) = 
c_1^{(i)}\bar n\cdot p_{K^*} \zeta_\perp^{BK^*}\\
&& \hspace{1cm} - 
m_B^2 \int_0^1 dz b_{1L}^{(i)}(z) \zeta_{J\perp}^{BK^*}(z) \nn 
\end{eqnarray}
Note that the leading order spectator operator does not contribute to the 
decay with a transverse vector meson in the final state.

The function $\zeta_{J\perp}^{BK^*}(z)$ appearing in the factorizable
term is defined as a convolution of the jet function with the 
light-cone wave functions of the $K^*$ and $B$ mesons
\begin{eqnarray}
\zeta_{J\perp }^{BK^*}\!(z) \!=\! \frac{f_B f_{K^*}^T}{m_B}\!\int \!
d\!x d k_+ J_\perp(x,z,k_+) \phi_B^+(k_+) \phi_{K^*}^\perp(x)
\end{eqnarray}
Using the result for the jet function Eq.~(\ref{jets}) at
leading order in $\alpha_s(\mu_c)$, the integrals can be 
performed explicitly, and the function $\zeta_{J\perp}(z)$
is given by
\begin{eqnarray}\label{zetaJtree}
\zeta_{J\perp }^{BK^*}(z) = \frac{\pi \alpha_s C_F}{N_c} \frac{f_B f_{K^*}^T}{m_B}
\frac{1}{\lambda_{B+}} \frac{\phi_{K^*}^\perp(z)}{\bar z}
\end{eqnarray}
with the first inverse moment of the B wave function 
\begin{eqnarray}
\lambda_{B+}^{-1} = \int_0^\infty dk_+ \frac{\phi_+^B(k_+)}{k_+}
\end{eqnarray}

The corresponding amplitudes for the charge conjugate mode
$B\to K^*_n\ell^+ \ell^-$ are obtained from this by exchanging
$H_+ \leftrightarrow H_-$.
The vanishing of the right-handed helicity amplitude at leading order in
$\Lambda/m_b$ is a general result for the soft (nonfactorizable) component
of the form factors in $B\to M_n$, combined with the absence of the
factorizable contribution for this particular transition.
This result is usually expressed as two exact symmetry relations among the
tensor and vector $B\to V$
form factors at large recoil \cite{ffrel,befe}.

We proceed next to discuss the multibody
decays $\bar B\to \bar K\pi \ell^+\ell^-$, in the kinematical
region with $q^2 \sim 4$ GeV$^2$.  According to the discussion
of Sec.~II.A, the form of the factorization relation is different
in the three regions of the Dalitz plot shown in Fig.~\ref{figdal}.
Our main interest is in the region I, with one energetic kaon, and 
a soft pion. In this paper we prove a new factorization relation for
the transverse helicity amplitudes in this region.

We consider for definiteness the mode $\bar B^0 \to K_n^- \pi^+ \ell^+\ell^-$.
Collecting the partial results in Sec.~II.A, 
we find the following results for the transverse helicity amplitudes in this
mode (with $i=V,A$), valid in the region I 
\begin{eqnarray}\label{HKplpimi}
&& H_-^{(i)}(\bar B^0 \to K_n^- \pi^+ \ell^+\ell^-) = 
H_{\rm nf}^{(i)} + \delta_{i,V} \frac23 H_{\rm sp}^{(u)} \\
&& H_+^{(i)}(\bar B^0 \to K_n^- \pi^+ \ell^+\ell^-) = H_{\rm f}^{(i)}
\end{eqnarray}
where the three terms correspond to the nonfactorizable, spectator and factorizable
terms in Eq.~(\ref{Hsum}), respectively. They are given by
\begin{eqnarray}\label{Hnf}
&& H_{\rm nf}^{(i)} = c_1^{(i)}(\bn\cdot p_K, \mu)  \zeta_\perp^{BK\pi} \,,\\
&& H_{\rm f}^{(i)} = -\frac12 f_K (\varepsilon_+^*\cdot p_\pi) \\
&& \times \int_0^1 dz dx b_{1R}^{(i)}(z) \int_{-p_\pi^+}^{\infty} dk_+ 
J_\parallel (x,z,k_+) S(k_+) \phi_K(x)\,, \nn \\
&& H_{\rm sp}^{(q)} = \frac{(4\pi)^2}{q^2} 
f_K (\bn\cdot p_K)(\varepsilon_-^*\cdot p_\pi) 
\label{Hsp}\\
&& \times \int_0^1 dx b_{\rm sp}^{(q)}(x) \phi_K(x) 
\int_{-p_{\pi-}}^{\infty} dk_- J_{\rm sp}(k_-)
S(k_-) \nn \,.
\end{eqnarray}

The nonfactorizable operators contribute only to the left-handed helicity
amplitudes, and are given by the soft functions $\zeta_\perp^{BK\pi}$.
They are the same for both $i=V,A$ amplitudes. Furthermore, the same
soft functions would appear also in factorization relations for semileptonic
decays into multibody states, such as $\bar B\to \pi_n \pi \ell\bar\nu$.
This universality is the analog of the form factor relations for the 
nonfactorizable amplitudes \cite{ffrel,befe}, well-known from one-body
decays, to the multibody case.

The factorizable operators give nonvanishing 
contributions $H_{\rm f}^{(i)}$ to the right-handed helicity amplitudes. The appearance
of these contributions is a new effect, specific to the multibody decays
\cite{pol,GPchiral}. On the other hand, the spectator operator contributes
only to the left-handed helicity amplitudes.

The helicity amplitudes for  all other 
$\bar B\to K_n\pi\ell^+\ell^-$ decays can be obtained in a similar way. 
The results are tabulated in Table \ref{Htable}.

\begin{table}
\caption{\label{Htable} 
The transverse helicity amplitudes $H_\pm^{(i)}$ with $i=V,A$,
for the different charge states in $\bar B\to \bar K_n\pi \ell^+\ell^-$ 
decays, at leading order in $\Lambda/Q$. The building blocks $H_{\rm nf}^{(i)},
H_{\rm f}^{(i)}, H_{\rm sp}^{(q)}$ are given in Eqs.~(\ref{Hnf})-(\ref{Hsp}).
}
\begin{ruledtabular}
\begin{tabular}{|r|c|c|}
 &  $H_{-}^{(i)}$ & $H_{+}^{(i)}$ \\
\hline
$\bar B^0 \to K^-\pi^+\ell^+\ell^-$ & 
  $H_{\rm nf}^{(i)}  + \delta_{i,V} \frac23 H_{\rm sp}^{(u)}$ & 
  $H_{\rm f}^{(i)}$ \\
$K_S\pi^0\ell^+\ell^-$ & 
  $\frac12 (H_{\rm nf}^{(i)} - \delta_{i,V} \frac13 H_{\rm sp}^{(d)})$ & 
  $\frac12 H_{\rm f}^{(i)}$ \\
\hline
$B^- \to K^-\pi^0\ell^+\ell^-$ & 
  $\frac{1}{\sqrt2}(H_{\rm nf}^{(i)}  + \delta_{i,V} \frac23 H_{\rm sp}^{(u)})$ &
  $\frac{1}{\sqrt2}H_{\rm f}^{(i)}$ \\
$K_S \pi^-\ell^+\ell^-$ & 
  $\frac{1}{\sqrt2}(H_{\rm nf}^{(i)} - \delta_{i,V} \frac13 H_{\rm sp}^{(d)})$ & 
  $\frac{1}{\sqrt2}H_{\rm f}^{(i)}$ \\
\end{tabular}
\end{ruledtabular}
\end{table}

The structure of these results displays universality of hard-collinear effects.
This is manifested as the fact that all factorizable terms $H_{\rm f}^{(i)}$ 
depend on the same $x,k_+$ convolution 
\begin{eqnarray}
I_{\rm f}(z,p_\pi) \equiv \int_0^1 dx
\int_{-p_{\pi}^+}^{\infty} dk_+ 
J_\parallel (x,z,k_+) S(k_+) \phi_K(x)
\end{eqnarray}
and all spectator contributions depend on the same $k_-$ integral
\begin{eqnarray}
I_{\rm sp}(n\cdot q) = 
\int_{-p_{\pi}^-}^{\infty} dk_- 
J_{\rm sp}(k_-) S(k_-) 
\end{eqnarray} 
This universality is similar to that appearing in other factorization relations
in exclusive decays. Examples are the relation
between the rare leptonic decays $B_s\to \ell^+\ell^-\gamma$ and the
radiative leptonic decay $B\to \gamma\ell\bar\nu$ \cite{LPW}, and the 
relation among factorizable contributions in heavy-light form factors at 
large recoil, and the nonleptonic $B$ decays into two light mesons \cite{bprs}. 
The integral $I_{\rm sp}(n\cdot q)$ appears also in the leading order factorization
relation for exclusive semileptonic radiative decay $B\to \pi\gamma 
\ell\bar \nu$ \cite{rad}, and could be determined from measurements
of this decay.

Treating the spectator amplitude using the $n\cdot q \Lambda/q^2$ expansion,
the amplitude $H_{\rm sp}^{(q)}$ can be obtained as explained from Eq.~(\ref{Hsp}),
at leading order in this expansion, by the substitution Eq.~(\ref{subst})
\begin{eqnarray}\label{Hsp1}
&& H_{\rm sp}^{(q)} = - \frac{4\pi^2}{q^2 \bn\cdot q} f_K m_B \bn\cdot p_K 
(\varepsilon_-^*\cdot p_\pi) f_T(t^2) \\
&& \qquad \times \int_0^1 dz b_{\rm sp}^{(q)}(z) \phi_K(z) \nn
\end{eqnarray}

In the numerical estimates of this paper, we will use the leading order
chiral perturbation theory result Eq.~(\ref{SChPT}) for the soft function
$S(k_+)$ in the resonance region
$k_+ > 0$. The contribution to the $k_+$ convolutions in $H_{\rm f}^{(i)}$
and $H_{\rm sp}^{(q)}$ from the transition region $-n\cdot p_\pi \leq k_+ \leq 0$ will be
neglected, which
can be expected to be a good approximation for very soft pions
(the region $E_\pi \leq 500 $ MeV, corresponding to the lower shaded region in 
Fig.~\ref{figdal}). 
We emphasize that, although
the use of the chiral perturbation theory result for $S(k_+)$ is restricted 
to part of the region (I), the factorization relations proved in this paper
are valid over the entire region (I). Their unrestricted application requires a model
for the soft function $S(k_+)$ whose validity extends beyond the limitations of
chiral perturbation theory.

With these approximations, additional universality emerges,
connecting the amplitudes in this problem to other B decays, to all orders in
the perturbative expansion at the hard-collinear scale.
The factorizable helicity amplitudes $H_{\rm f}^{(i)}$
take a simpler form, and can be written as
\begin{eqnarray}\label{HfChPT}
H_{\rm f}^{(i)} = \frac12 m_B^2 S_R(p_\pi)
 \int_0^1 dz b_{1R}^{(i)}(z) \zeta_{J}^{BK}(z)
\end{eqnarray}
where the nonperturbative dynamics is contained in the factorizable convolution
defined as
\begin{eqnarray}
\zeta_J^{BK}(z) \!=\! \frac{f_B f_K}{m_B}\!\int_0^1 \!\!\!\!
dx \!\int_0^{\infty} \!\!\!\! dk_+ \! J_\parallel(x,z,k_+) \phi_B^+(k_+) \phi_{K}(x)
\end{eqnarray}
The same function appears also in the factorizable contribution 
to the $B\to K$
form factors at large recoil \cite{bpsff}, and in the factorization
relation for nonleptonic 
decays $B\to KM$
with $M = \pi, K, \cdots$ a light meson \cite{bprs}.
The pion momentum dependence in Eq.~(\ref{HfChPT}) 
is contained in the function $S_R(p_\pi)$
given by \cite{GPchiral}
\begin{eqnarray}
S_R(p_\pi) = \frac{g}{f_\pi} \frac{\varepsilon_+^*\cdot p_\pi}
{v\cdot p_\pi + \Delta - i\Gamma_{B^*}/2}
\end{eqnarray}
with $\Delta = m_{B^*}-m_B \simeq 50$ MeV. 

A similar result is obtained for the spectator amplitude at leading
order in chiral perturbation theory, for which we find (treating the 
$q^2$ as a hard-collinear scale)
\begin{eqnarray}\label{HspChPT}
&& H_{\rm sp}^{(q)} = \frac{4\pi^2}{q^2} 
f_B f_K m_B (\bn\!\cdot\! p_K)
\Big(\frac{g}{f_\pi} \frac{\varepsilon_-^*\cdot p_\pi}{v\cdot p_\pi + \Delta}
\Big) \\
&& \qquad  \times
\int_0^1 dx b_{\rm sp}^{(q)}(x) \phi_K(x) 
\int_0^\infty dk_- J_{\rm sp}(k_-) \phi_+^B(k_-) \nn
\end{eqnarray}
The $k_-$ convolution in this relation is identical to that appearing 
in the leading order factorization relation for radiative semileptonic 
decays $B\to\gamma \ell \bar \nu$.

Adopting the approach of expanding in $n\cdot q \Lambda/q^2$, the 
result for $H_{\rm sp}^{(q)}$ requires the $B\to \pi$ form
factors, for which one finds at leading order in HHChPT \cite{wise,BuDo,TM,review}
\begin{eqnarray}
m_B f_T(t^2) = f_+(t^2) - f_-(t^2) = -g\frac{f_B m_B}{f_\pi} \frac{1}{E_\pi + \Delta}
\end{eqnarray}
We will use these expressions together with 
Eq.~(\ref{Hsp1}) in the numerical evaluations of Sec.~V.

\section{Decay rates and the FB asymmetry}

The differential decay rate for $B\to K\pi \ell^+\ell^-$ is given by
(see, e.g. \cite{KM})
\begin{eqnarray}
&& \frac{1}{\Gamma_0} \frac{d^2\Gamma}{dq^2 d\cos\theta_+ dM_{K\pi}^2 dE_\pi} = 
\frac{q^2}{2(4\pi)^3 m_B^2 m_b^5}\\
&&
\quad \times \Big\{2\sin^2\theta_+ (|H_0^V|^2 + |H_0^A|^2) \nn \\
&& +\, (1 + \cos^2\theta_+) (|H_+^V|^2 + |H_+^A|^2 + |H_-^V|^2 + |H_-^A|^2)\nn\\
&& \quad +\, 4\cos\theta_+  \mbox{Re}(H_-^V H_-^{A*} - H_+^V H_+^{A*})\Big\} \nn
\end{eqnarray}
with $\Gamma_0 = G_F^2\alpha^2/(32 \pi^4) |\lambda_t^{(s)}|^2 m_b^5$.
We denoted $\theta_+$
the angle between the direction of the positron momentum and the decay
axis in the rest frame of the lepton pair, for a fixed configuration of the 
hadronic state $K\pi$ defined by $(M_{K\pi}, E_\pi)$.

Integrating over $\cos\theta_+$ one finds for the forward-backward
asymmetry (FBA) defined as in Eq.~(\ref{AFBdef})
\begin{eqnarray}
A_{FB} \propto \mbox{Re }(H_-^V H_-^{A*} - H_+^V H_+^{A*})
\end{eqnarray}
This defines a triply differential asymmetry depending on $(q^2, M_{K\pi}, E_\pi)$.
Integrating also over $E_\pi$ gives a doubly differential $A_{FB}$ depending
only on $(q^2, M_{K\pi})$. We denote them with the same symbol, and distinguish
between them by their arguments.

The condition for a zero of the FBA can be written down straightforwardly
using the expressions for the helicity amplitudes in factorization given 
previously in Sec.~III. The equation for the zero is different in the
two regions (I) and (II), according to the different form of the factorization
relations in each of them. It is convenient to  write this equation in a 
common form in both regions, as
\begin{eqnarray}\label{eqzero}
\mbox{Re }\Big( c_1^{(V)} (M_{K\pi}, q^2) - a ) = 0\,.
\end{eqnarray}
The quantity $a$ stands for the contribution of the factorizable and
spectator type amplitudes, and in general is a function of all kinematical
variables $(M_{K\pi}, q^2, E_\pi)$.

In the region (II) with a collinear kaon and pion, this correction
is given by the (complex) quantity
\begin{eqnarray}
a_{\rm II} = 
-\frac{m_B^2}{E_{K^*}\,\bar n\cdot q}\,C_7^{\text{eff}}\,\frac{\zeta_{J\perp }^{BK^*\text{eff}}}{\zeta_\perp^{BK^*}}
\end{eqnarray}
where the factorizable coefficient, 
$\zeta_{J\perp }^{BK^*\text{eff}}$, which has implicit dependence on $(M_{K\pi},q^2)$, is defined by
\begin{eqnarray}
 \int_0^1 dz b_{1L}^{(V)}(z)  \zeta_{J\perp }^{BK^*}(z)= 
-\frac{2n\cdot q}{q^2}\,C_7^{\text{eff}} \,\zeta_{J\perp }^{BK^*\text{eff}} 
\end{eqnarray}
The zero of the FBA in region (II) was considered previously
in Refs.~\cite{fb,ABHH,fbcor1,fbcor2,fbcor3}, treating the problem as
a one-body decay $B\to K^*\ell^+\ell^-$.

In the region (I) with a soft pion and a collinear kaon, this correction
contains two terms, arising from the spectator and the factorizable 
contributions, respectively. We adopt everywhere in the following the
leading order chiral perturbation theory results for the amplitudes
given in Eqs.~(\ref{HfChPT}) and  (\ref{HspChPT}).
Working at tree level in matching at the
scale $\mu = Q$, but to all orders in the hard-collinear scale, one finds
\begin{eqnarray}\label{aI}
a_{\rm I} &=& - e_q S_R^*(p_\pi) \frac{h^{(q)}_{\rm sp}}{\zeta_\perp^{BK\pi}} \\
& &
+ \Big(\frac{m_B^2}{4E_K} \Big)^2 |S_R(p_\pi)|^2 C_9^{\rm eff} 
\frac{\zeta_J^{BK} \zeta_J^{BK eff}}{|\zeta_\perp^{BK\pi} |^2} \nn
\end{eqnarray}
The first term contains the contribution of the spectator amplitude,
and depends on the charge states of the final and initial state through
the superscript $q=u,d$, denoting the flavor of the quark attaching to the
photon. The quantity $h^{(q)}_{\rm sp}$, given by
\begin{eqnarray}\label{hspdefhc}
h^{(q)}_{\rm sp} &=& \frac{4\pi^2}{q^2}
f_B f_K m_B (\bn\!\cdot\! p_K)
\int_0^1 dx b_{\rm sp}^{(q)}(x) \phi_K(x) \\
& &\times 
\int_0^\infty dk_- J_{\rm sp}(k_-) \phi_+^B(k_-) \nn
\end{eqnarray}
depends on kinematic variables only through the explicit factor
  of $\bar n\cdot p_K/q^2$, to the order we are working.
For completeness, we quote also the expression for this amplitude at leading order
in $n\cdot q \Lambda/q^2$, which is used in the actual numerical computation of Sec.~V
\begin{eqnarray}\label{hspdef}
h^{(q)}_{\rm sp} = \frac{4\pi^2}{q^2 \bn\cdot q}
f_B f_K m_B (\bn\!\cdot\! p_K)
\end{eqnarray}
The first term in Eq.~(\ref{aI}) contributes to Re$(a_{\rm I})$ with an 
undetermined sign, depending on the unknown $\zeta_\perp^{BK\pi}$.

The second term in Eq.~(\ref{aI}) is due to the right-handed
helicity amplitudes $H_+^{(i)}$. Its dependence on $E_\pi$
is explicit in $S_R(p_\pi)$, and the remaining factors depend only on $(M_{K\pi},q^2)$. 
The factorizable coefficients in the numerator are defined as
\begin{eqnarray}
&& \int_0^1 dz b_{1R}^{(V)}(z) \zeta_{J}^{BK}(z) \equiv \frac{1}{\bn\cdot p_K}C_9^{\rm eff}
\zeta_J^{BK\rm eff}\\
&& \int_0^1 dz b_{1R}^{(A)}(z) \zeta_{J}^{BK}(z) \equiv \frac{1}{\bn\cdot p_K}C_{10}
\zeta_J^{BK}
\end{eqnarray}
where we kept only the tree level matching result for $b_{1R}^{(A)}(z)$. Numerical evaluation 
of these factorizable amplitudes in the next section shows that the contribution
of this term to Re$(a_{\rm I})$ is positive.

To explore the implications of these results, let us assume as a starting point that 
the helicity amplitudes $H_\pm$ in both regions
are dominated by the nonfactorizable contributions,  proportional to $\zeta_\perp^{BK\pi}$ (in region
(I)), and $\zeta_\perp^{BK^*}$ (in region (II)). 
This corresponds to taking $a=0$ in both regions.
In this approximation, the condition for the zero of the FBA reads 
simply Re($c_1^{(V)})=0$, which can be solved exactly.
For the $B\to M_n$ transition, this condition reproduces the
well-known result following from the large energy form factor relations
\cite{fb,ABHH,ffrel,fbcor1,fbcor2}. 

Since the zero of the FBA is related to the vanishing of the 
Wilson coefficient Re $(c_1^{(V)})$, 
such a zero must be present also for B decays into multibody
states containing one energetic kaon~\footnote{The existence of a zero of the
FBA in inclusive $B\to X_s \ell^+\ell^-$ decays was noticed and
studied in Refs.~\cite{FBAincl,FBAincl1}. In this paper we address only the
semi-inclusive decays $B\to K_n \pi$ with low hadronic invariant mass, 
for which the inclusive
methods of Refs.~\cite{FBAincl,FBAincl1} are not applicable.}.
In particular, the FBA in $B\to K_n \pi 
\ell^+ \ell^-$ must have a zero 
at a certain point $q_0^2=q_0^2(M_{K\pi})$
which depends only on the invariant mass of the hadronic system. 
Adding the second term in Eq.~(\ref{eqzero}) shifts the position
of this zero, and introduces a dependence on the pion energy, $q_0^2 =
q_0^2(M_{K\pi},E_\pi)$.  This extends the well-known result for the zero of the FBA in
$B\to K^* \ell^+ \ell^-$ to multibody hadronic states.

It is interesting to comment on
existing computations of the decay amplitudes and FB asymmetry in 
$B\to K\pi \ell^+\ell^-$ \cite{ABHH,fbcor1,fbcor2,fbcor3}, which keep only the $K^*$ 
resonant amplitude. Of course, this is justified in the region (II), where the 
pion and the kaon are
collinear. However, in the region (I) this contribution is in fact  parametrically suppressed,
since by leading order soft-collinear factorization, the $K^*_n K_n \pi_S$ 
vertex does not exist at leading order in $O(\Lambda/m_b)$.

Computing the factorizable corrections in the region (I) parameterized by the 
term $a_{\rm I}$ requires
that we know the nonfactorizable soft function $\zeta_\perp^{BK\pi}$. 
There are several possible ways of determining $\zeta_\perp^{BK\pi}$ from data.
For example, according to Eq.~(\ref{HKplpimi}), the helicity amplitude $|H_-^A|$ receives no
factorizable or spectator contributions. Assuming that it can be isolated, its
measurement would give a clean determination of $|\zeta_\perp^{BK\pi}|$.
Another method involves measuring $H_-^{(V)}$ in decays with a neutral kaon in the
final state, for which the spectator contribution is small (see Table I).
We assume that the factorizable coefficients can be computed in perturbation theory.
We postpone a detailed numerical analysis for Sec.~V, and discuss in the following
general properties of the zero of the FBA which are independent on the details of the
hadronic parameters.

\begin{figure}
\hskip-20pt\includegraphics[height=5.3cm]{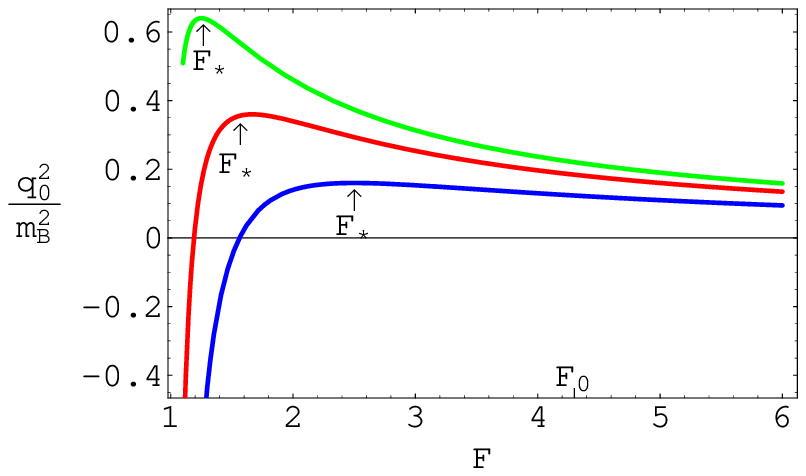}
\caption{\label{fig:qualitative} Plot of the zero in the
  Forward-Backward Asymmetry as a function of the parameter $F$ of
  Eq.~(\ref{FBAs2}), for several values of $M_{K\pi}$. The top (green) curve has
  lowest $M_{K\pi}$ and the bottom (blue) one has highest  $M_{K\pi}$.
  $F_0$ corresponds to the value of $F$ in the absence of factorizable
  and spectator corrections. The maxima of the curves, labelled by
  $F_*$, lie to the left of $F_0$ for all relevant values of $M_{K\pi}$. }
\end{figure}

\subsection{Qualitative discussion of the zero of the FBA}
\label{sec:qual}
Before proceeding with the details of the numerical study, we would like
to discuss some of the qualitative properties of the zero of the FBA in the
multibody decay $B\to K\pi\ell^+\ell^-$. The general behaviour of the
solution can be seen by studying the solutions of the simplified equation
\begin{eqnarray}\label{FBAsimple}
C_9 + 2 m_b \frac{n\cdot q}{q^2} C_7 - a(M_{K\pi}) = 0
\end{eqnarray}
where 
$a(M_{K\pi})$ denotes the factorizable correction.
In this simplified version of Eq.~(\ref{eqzero}) we have neglected 
additional dependence of $a$ on $q^2$, which is adequate if one
is interested in the qualitative change in the zero in the FBA
at fixed $q^2$. We also have
neglected here the radiative corrections. They introduce small
logarithmic dependence on $M_{K\pi}$, and do not change the
qualitative features of the solution.

It is convenient to write the equation Eq.~(\ref{FBAsimple}) in an equivalent form
\begin{eqnarray}\label{FBAs2}
\frac{n\cdot q}{q^2} = - \frac{1}{2m_b C_7} (C_9 - a(M_{K\pi})) \equiv \frac{F(M_{K\pi})}{m_B}
\end{eqnarray}
The solution to this equation gives the location, $q^2_0$, of
  the zero of the FBA:
\begin{equation}
\label{eq:q2zero-qual}
q^2_0(M_{K\pi})=\frac{m_B^2}{F}-\frac{M^2_{K\pi}}{F-1}
\end{equation}
 The condition that the FBA-zero lies in the physical region, $0\leq
q_0^2\leq(m_B-M_{K\pi})^2$, imposes constraints on $F=F(M_{K\pi})$. The upper
bound $q_0^2\leq(m_B-M_{K\pi})^2$ implies that $F(M_{K\pi})\geq 1$, while
from $0\leq q_0^2$ we learn that
\begin{equation}
\label{eq:F-cond}
F(M_{K\pi})\geq1+\frac{M^2_{K\pi}}{m_B^2-M_{K\pi}^2}\geq1\,.
\end{equation}
In terms of the correction $a$ the condition for the existence of a
FBA-zero is therefore
\begin{equation}
a(M_{K\pi})\leq C_9+\frac{2m_bm_B}{m_B^2-M_{K\pi}^2}C_7\,.
\end{equation}

If the function $F=F(M_{K\pi})$ is roughly constant and satisfies
the condition in (\ref{eq:F-cond}), then Eq.~(\ref{eq:q2zero-qual})
gives that the zero of the FBA decreases with increasing
$M_{K\pi}$. These conditions hold if $a(M_{K\pi})\ll C_9$. Since we expect
the correction term to be small we also expect the FBA-zero to
decrease as $M_{K\pi}$ increases. 

We can gain further insights into the solution by considering $q_0^2$
as a function of $F$ for fixed $M_{K\pi}$ in
Eq.~(\ref{eq:q2zero-qual}). Fig.~\ref{fig:qualitative} shows plots of
$q_0^2$ vs.\ $F$ for several fixed values of $M_{K\pi}$. The sequence of
curves moves down with increasing $M_{K\pi}$, which just restates the
observation that at fixed $F$ the zero decreases with increasing
$M_{K\pi}$. The point $F=F_0$ corresponds to $a=0$. The maxima of the
curves are at $F_*(M_{K\pi})=(1-M_{K\pi}/m_B)^{-1}$, and, for physical
values they lie to the left of $F_0$, that is, $F_*<F_0$. To see how
$q_0^2$ depends on $M_{K\pi}$ when $F=F(M_{K\pi})$ is not constant,
consider as starting value a point on the top curve. An increase in
$M_{K\pi}$ first moves the point down to a lower curve (as if $F$ were
constant), and then also along the lower curve to a different value of
$F$. The region to the right of $F_*$ is most interesting since we
expect the physical function to lie in a region of $F$ close to
$F_0$. In this region, if $F$ increases with $M_{K\pi}$ then $q_0^2$
decreases with (increasing) $M_{K\pi}$. The opposite is not necesarily
true: whether $q_0^2$ increases with $M_{K\pi}$ or not depends on how
steeply $F$ decreases with $M_{K\pi}$.

Next, we would like to understand what the effect of changing $a\to a +\delta
a$ is, corresponding to adding correction terms sequentially. A shift
$\delta F>0$ for fixed $M_{K\pi}$ corresponds to moving to the right along a
fixed curve. In the region to the right of $F_*$ this decreases
$q_0^2$. Note that $\delta a =(2m_bC_7/m_BC_9)\delta F$ and
$C_7/C_9<0$. Therefore, an 
increase in  $a$ gives an increase in $q_0^2$.

\section{Numerical study}

We investigate in this section the numerical effects of the new right-handed 
amplitude on the position of the zero of the FB asymmetry. As explained, 
we treat separately the decay amplitudes in the two regions (I) and (II), and
ignore the contribution from region (III). For definiteness, we consider here the
mode $\bar B^0 \to  K^-\pi^+\ell^+\ell^-$ for which the soft pion detection efficiency
is better than for the neutral pion modes.

The decay amplitudes in the collinear region (II) will be represented by a
Breit-Wigner model, as
\begin{eqnarray}
&& H_-^V(\bar B^0 \to K^- \pi^+ \ell^+\ell^-) = h_-^V(\bar B\to \bar K^*) \\
&& \qquad \times g_{K^* K\pi} (\varepsilon_-^*\cdot p_\pi) BW_{K^*}(M_{K\pi})\nn \\
&& H_-^A(\bar B^0 \to K^- \pi^+ \ell^+\ell^-) = h_-^A(\bar B\to \bar K^*) \\
&& \qquad \times g_{K^* K\pi} (\varepsilon_-^*\cdot p_\pi) BW_{K^*}(M_{K\pi})\nn 
\end{eqnarray}
and $H_+^{V,A} = 0$, with
$h_-^{V,A}$ the one-body helicity amplitudes for $B\to K^* \ell^+\ell^-$ 
given above in Eq.~(\ref{H-I}).
The Breit-Wigner function corresponding to a $K^*$ resonance is defined as
\begin{eqnarray}
BW_{K^*}(M) = \frac{1}{M^2 - M_{K^*}^2 + i M_{K^*} \Gamma_{K^*}}
\end{eqnarray}
Finally, the $K^{*0} K^+\pi^-$ coupling with a charged pion can be determined from the
total $K^* \to K\pi$ width, $\Gamma = g_{K^*K\pi}^2 p_\pi^3/(16\pi m_{K^*}^2)$,
with the result $g_{K^*K\pi} = 9.1$.

\begin{figure}
\includegraphics[height=6cm]{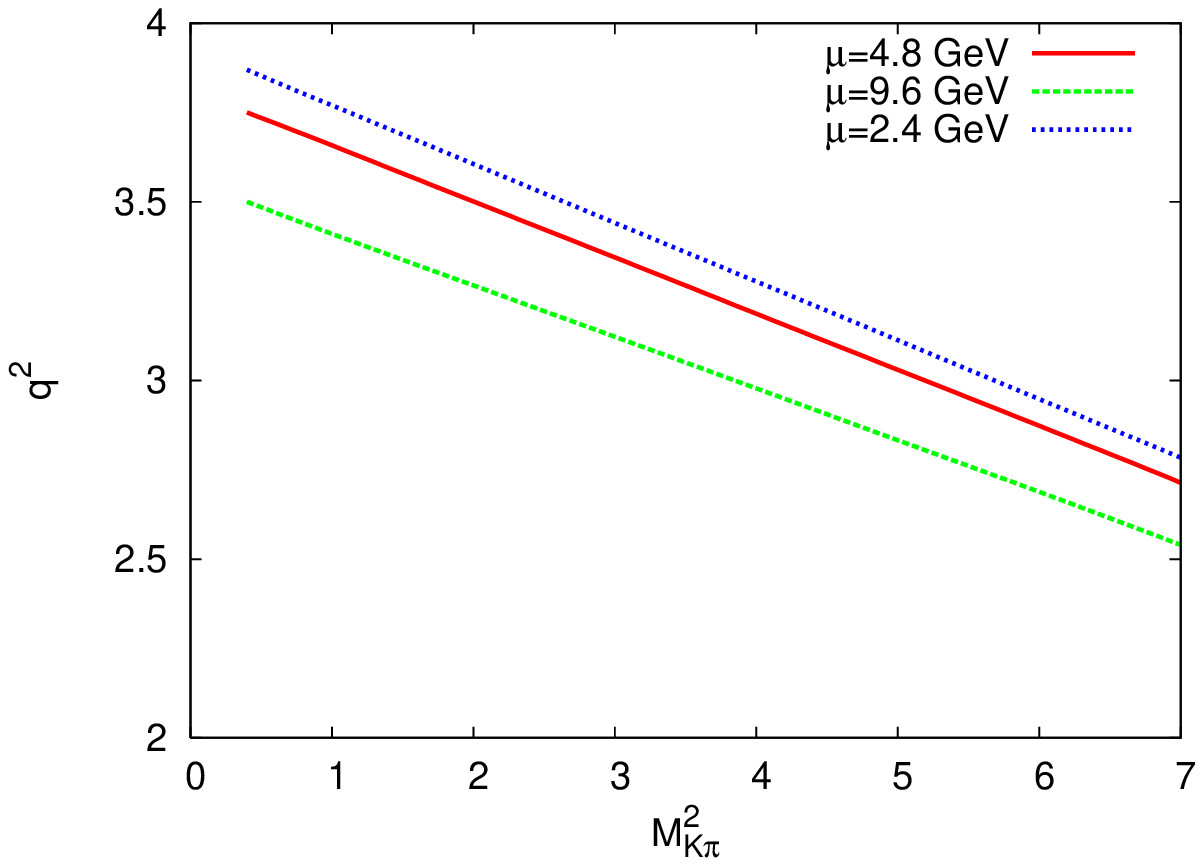}
\caption{\label{fig3} Plot of the position of the zero of the
forward-backward asymmetry
$q_0^2=q_0^2(M_X)$ as a function of the invariant mass of the $K\pi$
system, obtained by neglecting the factorizable contributions to the
helicity amplitudes, for different values of the renomalization point
$\mu$.}
\end{figure}

The factorization relations in region (I) require the nonfactorizable
amplitude $\zeta_\perp^{BK\pi}$. 
In the absence of experimental information about this quantity, we adopt
a $K^*$ resonance model for it, defined as 
\begin{eqnarray}
&& \zeta_\perp^{BK\pi}(M_{K\pi}, E_\pi) = \bn\cdot p_{K^*} \zeta_\perp^{BK^*} \\
&& \qquad \times g_{K^*K\pi}
(\varepsilon_-^*\cdot p_\pi) BW_{K^*}(M_{K\pi})\,.\nn
\end{eqnarray}
where the kinematical factor $\bn\cdot p_{K^*} = 2E_{K^*} = \frac{1}{m_B}(m_B^2-q^2)$
can be chosen corresponding to an on-shell $K^*$ meson.
For the kinematical dependence of the soft function $\zeta_\perp^{BK^*}(q^2)$
we adopt a modified pole shape \cite{fbcor3,BZ}
\begin{eqnarray}\label{zmodel}
\zeta_\perp^{BK^*}(q^2) = \frac{\zeta_\perp^{BK^*}(0)}{1 - 1.55 \frac{q^2}{m_B^2} +
0.575 (\frac{q^2}{m_B^2})^2}
\end{eqnarray}
and quote results corresponding to the two values $\zeta_\perp^{BK^*}(0) = 0.3$ 
and $0.1$. These two choices should cover both cases of soft-dominated, and 
hard-dominated tensor form factor.

We will use this model to define a FBA differential in $(q^2, M_{K\pi})$, integrated
over the pion energy $E_\pi$. Separating the contributions from the regions (I) and 
(II), this is given by
\begin{eqnarray}
A_{\rm FB}(q^2, M_{K\pi}) &=& \int_{\rm (I)} dE_\pi \mbox{Re }[H_-^V H_-^{A*} - H_+^V H_+^{A*}] \nn\\
& & + \int_{\rm (II)} dE_\pi \mbox{Re }[H_-^V H_-^{A*}]
\end{eqnarray}
The integration over $E_\pi$ can be simplified by approximating
$\bar n\cdot p_K\approx m_B-\bar n\cdot q$ in region I. Then the result can be
expressed in terms of three phase space 
integrals $I_{0,1,2}$, arising from region (I), 
and another integral $\bar I_0$ in region (II), defined as
\begin{eqnarray}
&& I_j = \int_{E_\pi^{\rm min}}^{E_\pi^{\rm cut}} dE_\pi \frac{|\varepsilon_+\cdot p_\pi|^2}
{(E_\pi + \Delta)^j}\,,\,\, j = 0,1,2\\
&& \bar I_0 = \int_{E_\pi^{\rm cut}}^{E_\pi^{\rm max}} dE_\pi  |\varepsilon_+\cdot p_\pi|^2
\,.
\end{eqnarray}
These integrals depend implicitly on $(M_{K\pi},q^2)$. Numerically, for
$E_\pi^{\rm cut}=500~\text{MeV}$ we find 
$I_0=0.019~\text{GeV}^3$, $I_1=0.047~\text{GeV}^2$,
$I_2=0.12~\text{GeV}$ and 
$\bar I_0=0.12~\text{GeV}^3$, at $M_{K\pi}=1~\text{GeV}$ and $q^2=4~\text{GeV}^2$.

\begin{widetext}
The zero of the FBA
is given by the solution of the equation
\begin{eqnarray}
\label{eq:qzero2}
\mbox{Re } [c_1^{(V)} - a_{\rm sp} - a_{\rm f} ] = 0\,,
\end{eqnarray}
where $a_{\rm sp}$ and $ a_{\rm f}$ denote the factorizable contributions, arising
from the spectator amplitude, and from the factorizable (both in one-body and in the
two-body amplitudes). The $a_{\rm sp}$ coefficient is
given by
\begin{equation}
\label{asp-model}
a_{\rm sp} = -\frac23 \frac{g}{f_\pi} \frac{h_{\rm sp}^{(u)}}{\bn\cdot p_{K^*} 
g_{K^*} \zeta_\perp^{BK^*}}
\frac{I_1}{I_0 + \bar I_0} [M_{K\pi}^2 - M_{K^*}^2 + iM_{K^*} \Gamma_{K^*}]
\end{equation}
where $h_{\rm sp}^{(q)}$ is defined in Eq.~(\ref{hspdef}).

The factorizable term contains contributions from the one-body
decay amplitude, and from the new right-handed amplitude appearing in the
two-body mode
\begin{equation}
\label{af-model}
a_{\rm f} = -\frac{2m_B^2}{\bn\cdot p_{K^*}\bn \cdot q}\, C_7^{\text{eff}}\,
 \frac{\zeta_{J\perp}^{BK^*\text{eff}}}{ \zeta_\perp^{BK^*}}  \frac{\bar I_0}{I_0 + \bar I_0} 
 + \frac{g^2 m_B^4}{4f_\pi^2 g^2_{K^*K\pi}(\bn\cdot p_K)^2(\bn\cdot p_{K^*})^2} 
C_9^{\rm eff}\frac{\zeta_J^{BK} \zeta_J^{BK\rm eff}}{(\zeta_\perp^{BK^*})^2} (M_{K\pi}^2 - M_{K^*}^2)^2
\frac{I_2}{I_0 + \bar I_0} 
\end{equation}
\end{widetext}

We compute the factorizable matrix elements using the leading order
jet functions from Eq.~(\ref{jets}). This gives for the integrals of the 
factorizable functions $\zeta_J(z)$
\begin{eqnarray}
\zeta^{BK}_J&=\frac{\pi\alpha_sC_F}{N_c}\frac{f_Bf_K}{m_B} \frac1{\lambda_{B+}}\int_0^1dz\,\frac{\phi_K(z)}{1-z}\,,\\
\zeta^{BK^*}_{J\perp}&=\frac{\pi\alpha_sC_F}{N_c}
\frac{f_Bf^T_{K^*}}{m_B} \frac1{\lambda_{B+}}\int_0^1dz\,\frac{\phi^T_{K^*}(z)}{1-z}\,,
\end{eqnarray}
and for the effective ones
\begin{eqnarray}
\zeta^{BK\text{eff}}_J&=&\zeta^{BK}_J\left[\left(1+\frac{2m_b(\mu)\bn\cdot q}{q^2}
\frac{C_7^{\text{eff}}}{C_9^{\text{eff}}}\right)\right.\nn\\
&&\qquad\qquad +\left.
\frac{e_uE_K^2}{m_b\bar n\cdot q}\frac{\bar C_2}{C_9^{\text{eff}}}I^{BK}_J\right]\,,\\
\zeta^{BK^*\text{eff}}_{J\perp}&=& \zeta^{BK^*}_{J\perp}\left[1+
\frac{2e_uE_{K^*}}{8m_b}\frac{\bar C_2}{C_7^{\text{eff}}}I^{BK^*}_{J\perp}\right]\,,
\end{eqnarray}
where
\begin{eqnarray}
I^{BK}_J&=\int_0^1dz\, t_\perp(z,m_c)\phi_K(z)\Big/ \int_0^1dz\frac{\phi_K(z)}{1-z}\,,\\
I^{BK^*}_{J\perp}&=\int_0^1dz\, t_\perp(z,m_c)\phi^T_{K^*}(z)\Big/ 
\int_0^1dz\frac{\phi^T_{K^*}(z)}{1-z}\,,
\end{eqnarray}
For the computation of the integrals
we use the $K^{(*)}$ light-cone wave functions 
\begin{eqnarray}
\phi_K(x) = 6x\bar x \Big(1 + 3 a_{1K} (2x-1) + {\frac32} a_{2K} [5(2x-1)^2-1] \Big)\nn\\
\end{eqnarray}
and analogous for $\phi_{K^*}^\perp(x)$, with coefficients $a_{iK^*}^\perp$. 
The values of the first two Gegenbauer moments are given in
Table~\ref{tablein}.  Also listed there are the remaining 
hadronic parameters used in the computation. 

The resulting values of the factorizable matrix elements are tabulated
in Table~\ref{tableout}, which also lists the effective Wilson
coefficients $C^{\text{eff}}_{7,9}$ (computed at $\mu=4.8$~GeV,
$q^2=4$~GeV$^2$). The effective matrix elements $\zeta^{BK\text{eff}}_J$
and $\zeta^{BK^*\text{eff}}_{J\perp}$ depend on $(M_{K\pi}, q^2)$, partly
through implicit dependence of the integrals $I^{BK}_J$ and
$I^{BK^*}_{J\perp}$. To gain some understanding of the relative importance
of the terms that contribute to the effective matrix elements, we
quote these integrals at $M_{K\pi}=m_{K^*},q^2=4.0\text{~GeV}^2$:
$I^{BK}_J=-0.704- 2.564i$ and $I^{BK^*}_{J\perp}=-0.566- 2.67i$.

\begin{table}[h!]
\caption{\label{tablein} Input parameters used in the numerical computation.}
\begin{ruledtabular}
\begin{tabular}{cc|cc}
$m_b^{\rm 1S}$ & $4.68 \pm 0.03$ GeV \cite{mb}   &   $f_K$ & 160 MeV \\
$\bar m_c(\bar m_c)$ & $1224 \pm 57$ MeV \cite{mc}  &  $f_{K^*}^T$   & 175 MeV  \\
$\alpha_s(M_Z)$ & $0.119$ & $a_{1K}$   & $0.3$  \\
$\lambda_{B+}$ & $350$ MeV &  $a_{2K}$  & $0.1$  \\
$g$ & $0.5$ \cite{ff} & $a_{1K^*}^\perp $   & $0.2$  \\
$f_B$ & 200 MeV &  $a_{2K^*}^\perp$  & $0.1$  \\
$\lambda^{(s)}_u/ \lambda^{(s)}_t$ & $-0.0106+0.0174i$~\cite{Charles:2004jd} & $g_{K^*K\pi} $ & 9.1 \\
\end{tabular}
\end{ruledtabular}
\end{table}

\begin{figure*}
\begin{tabular}{cc}
{\includegraphics[height=6cm]{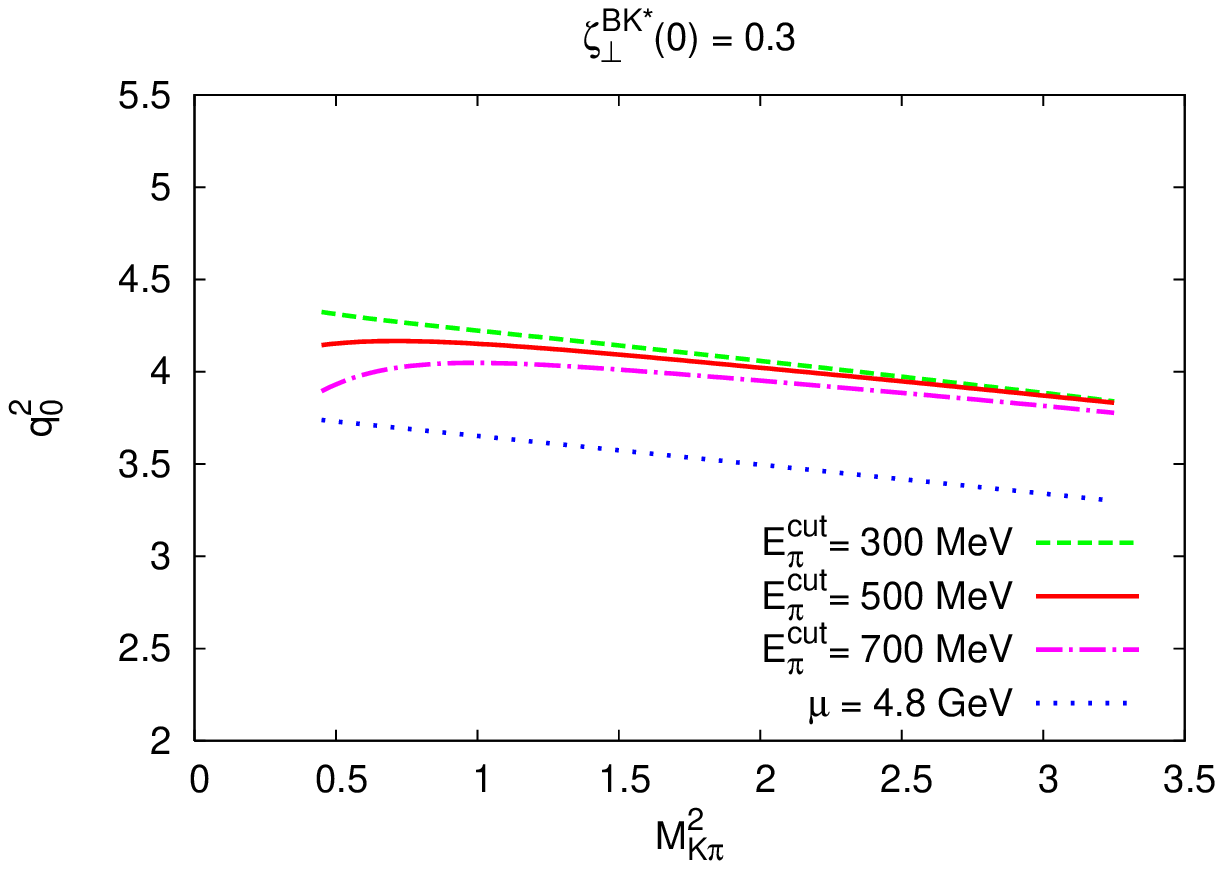}\hspace{1cm}
\includegraphics[height=6cm]{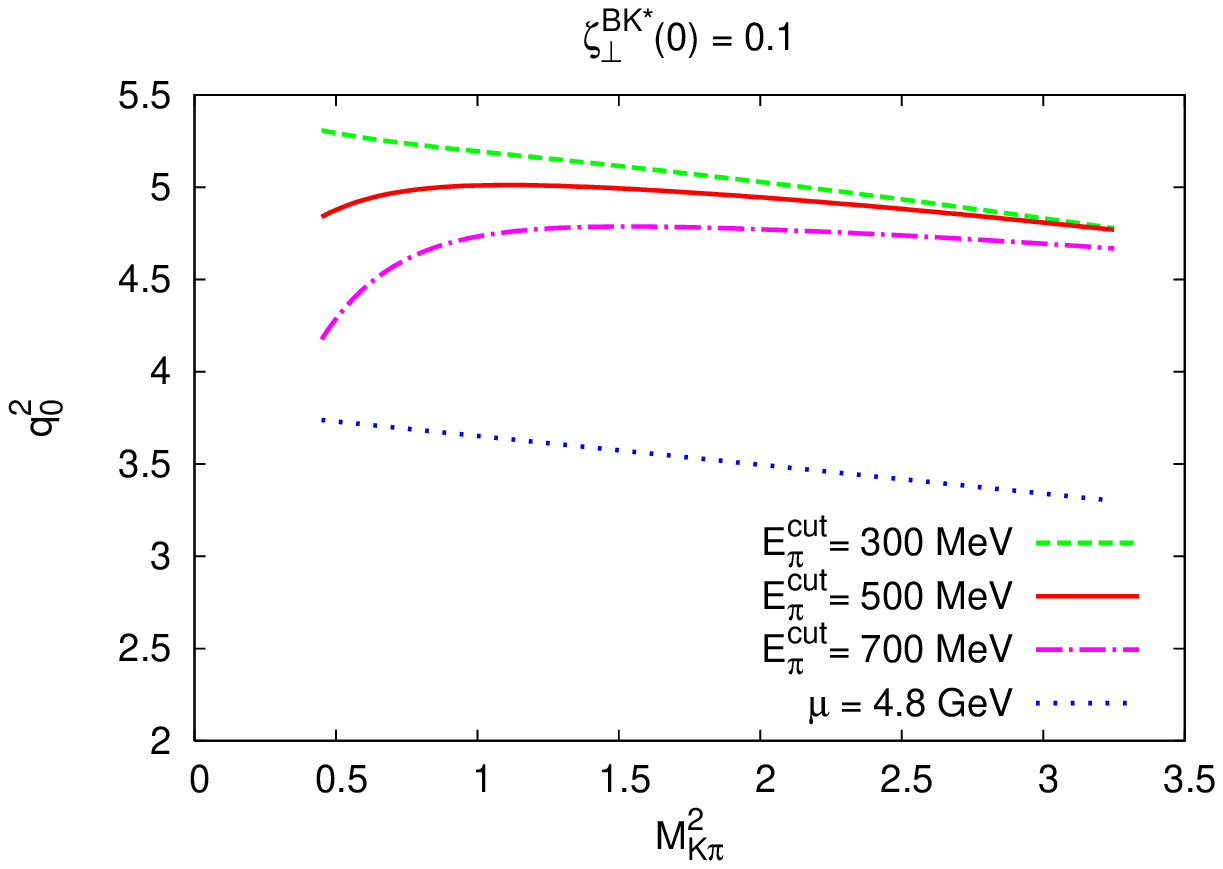}}
\end{tabular}
\caption{\label{fig7} Plot of the position of the zero of the
forward-backward asymmetry $q_0^2=q_0^2(M_{K\pi})$ as a function of the
invariant mass of the $K\pi$ system. The plots show the change in the
position of the zero due to the spectator and factorizable
amplitudes, Eq.~(\ref{eqzero}), for three values of the pion energy
cut-off $E_\pi^{\rm cut}=300, 500$~MeV and $E_\pi^{\rm cut}=700$~MeV
separating regions I and II. The dotted (blue) line denotes
the position of the zero in the absence of the factorizable and
spectator contributions (for $\mu = 4.8$ GeV).
The non-factorizable matrix element is taken to be 
$\zeta_\perp^{BK^*}(0) =
0.3$ (left) and $\zeta_\perp^{BK^*}(0) = 0.1$ (right).}
\end{figure*}

It is straightforward to estimate the numerical importance of
each term in the correction $a$ appearing in the equation Eq.~(\ref{eqzero})
for the zero of the FBA. To this end we evaluate at
$M_{K\pi}=m_{K^*}, q^2=4.0\text{~GeV}^2$, and use
$E^{\text{cut}}_\pi=500~\text{MeV}$ and $\zeta^{BK^*}_{\perp}(0)=0.3$ 
in the model of Eq.~(\ref{zmodel}). We obtain (all in GeV units):
\begin{eqnarray}
a_{\text{sp}} &\sim &  -(0.016+0.014i) (M_{K\pi} 
^2 - M_{K^*}^2 + iM_{K^*} \Gamma_{K^*})\,,\nn\\
\label{numestimate}
a_{\text{f}}  &\sim & (0.533 + 0.321 i) \\
&&\qquad\qquad + (0.004 -0.003i) (M_{K\pi}^2 - M_{K^*}^2)^2\,. \nn
\end{eqnarray}
We have kept explicit the rapidly varying dependence on the inverse
Breit-Wigner function, so we may get some idea of the relative
size of the coefficients. The first term in the factorizable amplitude
can be regarded as a negative correction to Re$(C_9^{\text{eff}})$ in 
$c_1^{(V)}$ of about $\sim 10\%$. Thus it 
effectively shifts the zero of the FBA upwards by the same amount. 
The second term in $a_{\text{f}}$  
and $a_{\text{sp}}$ are negligible on resonance but may be important at
large $M_{K\pi}$.

Using this model, we solve for the zeros of the FBA in $B\to K\pi \ell^+\ell^-$
decays, finding $q_0^2$ for given $(M_{K\pi}, E_\pi^{\rm cut})$. The
results are shown in Figs.~\ref{fig3}--\ref{fig7}. In Fig.~\ref{fig3}
we plot the result for $q_0^2(M_{K\pi})$ obtained by neglecting the
factorizable and spectator terms (the solution to $\mbox{Re
}(c_1^{(V)})=0$) for three values of the renormalization scale, $\mu =
2.4, 4.8, 9.6$~GeV. We used here NNLL results for the Wilson coefficients
and the 2-loop matrix elements of the operators $O_{1,2}$ obtained in
Ref.~\cite{Asatrian}.  The position of the zero at threshold is
\begin{eqnarray}
q_0^2|_{M_{K\pi}=M_K+M_\pi} = 3.75^{+0.12}_{-0.25} \mbox{ GeV}^2
\end{eqnarray}
where the uncertainty includes only the scale dependence. This result
depends only mildly on $M_{K\pi}$, as seen from Fig.~\ref{fig3}.

In Fig.~\ref{fig7} we show also the effect of including the
factorizable and spectator terms in Eq.~(\ref{eq:qzero2}), for three
values of the cut-off on the pion energy $E_\pi^{\rm cut}=300$ MeV, 500 MeV and
700 MeV separating regions I and II.
The parameters used in evaluating this plot are listed in
Tables~\ref{tablein} and~\ref{tableout}. 
For comparison, we present in Fig.~\ref{fig7}
results with $\zeta_\perp^{BK^*}(0)=0.3$ (left) and $\zeta_\perp^{BK^*}(0)=0.1$ 
(right). 
The latter choice effectively amplifies the
factorizable and spectator corrections, and can be taken as a
conservative upper bound of these effects. The dependence of the
results on $E_\pi^{\text{cut}}$ is an artifact of the separation of
regions discussed above and indicates the uncerainty in this
procedure. For this reason we have taken rather extreme values of
$E_\pi^{\text{cut}}$. The overall trend of $q_0^2(M_{K\pi})$ decreasing
towards the right of the plots is readily understood from the
qualitative discussion in Sec.~\ref{sec:qual}: it follows from the
correction term $a$ being small. Similarly, that the inclusion of
spectator and factorizable corrections tends to increase the value of
$q_0^2$ for fixed $M_{K\pi}$ follows from positivity of $a$.  

The results show a marked dependence (especially for small $M_{K\pi}$) of 
the zero position on the
pion energy cut-off $E_\pi^{\rm cut}$ which separates the regions (I) and (II).
This is essentially due to the dominance of the factorizable contribution in
region (II). A conservative way to use our results is to take the
smaller value of $E_\pi^{\rm cut}=300$ MeV, for which the chiral perturbation theory
result can be expected to be the most precise.

\begin{table}[h!]
\caption{\label{tableout} 
Results for the effective Wilson coefficients and factorizable and
spectator  matrix 
elements. The values of the
effective Wilson coefficients are at the scale $\mu = 4.8$ GeV and $q^2=4$ 
GeV$^2$. The factorizable matrix elements are computed at the scale
$\mu_c = 1.5$ GeV.}
\begin{ruledtabular}
\begin{tabular}{cc|cc}
 $[C_9^{\rm eff}]_{\rm NNLL}$ & $4.579+0.082i$ & $\zeta_J^{BK}$   & $0.036$  \\
$[C_7^{\rm eff}]_{\rm NNLL}$ & $-0.388-0.020i$ & $\zeta_{J\perp}^{BK^*}$  &
  $0.035$  \\
 & & $\frac{q^2 \bn\cdot q}{\bar n\cdot p_K}h^{(u)}_{sp} $ & $(0.135+0.124 i)\text{GeV}^3$\\
\end{tabular}
\end{ruledtabular}
\end{table}

So far our considerations were restricted to the case of 
$\bar B^0 \to K^-\pi^+\ell^+\ell^-$ decays. Going over to the CP conjugate mode
$B^0 \to K^+\pi^-\ell^+\ell^-$, the position of the zero could change because of 
direct CP violation present in the spectator amplitude $H_{\rm sp}^{(q)}$,
which is furthermore enhanced by a $4\pi^2$ factor (see Eq.~(\ref{Hsp})). Denoting
the factorizable corrections analogous to $a_{\rm sp}$ and $a_{\rm f}$ for the CP conjugate 
mode with $\bar a_i$, we find
\begin{eqnarray}
\bar a_{\text{sp}}\!\sim\!   -(0.016-0.014i) 
(M_{K\pi}^2\! -\! M_{K^*}^2\! +\! iM_{K^*} \Gamma_{K^*})\,,
\end{eqnarray}
such that the CP asymmetry in the position of the zero is induced through the finite
$K^*$ width, and is small. Beyond tree level, such an effect will be
introduced at order $\alpha_s(Q)$ through matching corrections to $b_{\rm sp}$.

We consider next another observable, the slope of the curve for the zero of the 
FBA in Fig.~\ref{fig7}.
From Eq.~(\ref{eq:q2zero-qual}), this is given by
\begin{eqnarray}\label{slope}
\frac{dq_0^2(M_{K\pi}^2)}{dM_{K\pi}^2} \!=\! - \frac{1}{F-1}\! +\!
\Big( \frac{M_{K\pi}^2}{(F-1)^2} - \frac{m_B^2}{F^2} \Big) \frac{dF(M_{K\pi}^2)}{dM_{K\pi}^2}
\end{eqnarray}
$F(M_{K\pi}^2)$ is defined in Eq.~(\ref{FBAs2}) and depends on the Wilson 
coefficients $C_{7,9}$, also on the factorizable contributions  $a(M_{K\pi})$. 
The last term contributes through the $M_{K\pi}$ dependence of  $a(M_{K\pi})$,
and is 
\begin{eqnarray}
\frac{dF(M_{K\pi}^2)}{dM_{K\pi}^2} = \frac{m_B}{2m_b C_7} 
\frac{da(M_{K\pi}^2)}{dM_{K\pi}^2} \simeq 0.02\,,
\end{eqnarray}
where we used the result Eq.~(\ref{numestimate}) for $a_{\rm sp}(M_{K\pi})$ and 
neglected the tiny contribution from $a_f$. The contribution of this term to 
Eq.~(\ref{slope})
is multiplied with a factor of order 1-2. Thus, even assigning this
estimate a conservative error of $\sim 200\%$, its contribution to the slope 
Eq.~(\ref{slope}) for values $M_{K\pi} \sim 1 $ GeV, is negligible compared to the 
first term depending only on $F$ (recall that $F \sim 4$). 
This is also seen in the curves in Fig.~\ref{fig7}, whose slopes are essentially
the same for all choices of the hadronic parameters considered. This shows that a
measurement of the slope of the zero could provide a useful source of information 
about
the Wilson coefficients $C_{7,9}$, but without the hadronic uncertainties associated with
the absolute position of the zero.

\section{Conclusions}  

We studied in this paper the helicity
structure of the exclusive rare $\bar B\to \bar K \pi \ell^+\ell^-$ decays
in the region of phase space with one energetic kaon and a soft pion.
In this region the helicity amplitudes are given by new factorization relations,
containing an universal soft matrix element, and
a new nonperturbative matrix element for the $B\to \pi$ transition
analogous to the off-forward parton distribution functions.

The most important difference with the $\bar B\to \bar K^*_n \ell^+\ell^-$
decays at large recoil is the appearance in the multibody case of a
nonvanishing right-handed helicity amplitude $\bar B\to [\bar K
\pi]_{h=+1}\ell^+\ell^-$ at leading order in $\Lambda/m_b$.  
This can be computed in factorization, in terms of the $B\to \pi$
off-forward matrix element of a nonlocal heavy-to-light operator. 
In the soft pion limit this nonperturbative matrix element can be computed
in chiral perturbation theory, and is related to the B meson light-cone wave function
\cite{GPchiral}.

We explored the implications of these results for the existence of a
zero of the forward-backward asymmetry of the lepton momentum,
pointing out two new results.  First, the FBA has a zero also for
nonresonant $B\to K \pi \ell^+\ell^-$ decays, occuring at a determined
value of the dilepton invariant mass $q_0^2(M_{K\pi})$, depending on the 
hadronic invariant mass $M_{K \pi}$. Second, there are calculable 
corrections to the position of the zero, which can be computed in factorization.
We use the factorization relations derived in this paper to compute
these correction terms. We present explicit numerical results working
at leading order in chiral perturbation theory \cite{GPchiral}, and show that
the results for the zero of the FBA in $\bar B\to [\bar K\pi]_{K^*}\ell^+\ell^-$ 
hold to a good precision also in the nonresonant region.
\vspace{0.5cm}

\appendix*
\section{Effective Wilson coefficients}

We collect here for convenience the expressions for 
the effective Wilson coefficients used in the numerical study of 
Section IV.
Working to NNLL order, they are given by
\begin{eqnarray}
&& C_9^{\rm eff} = C_9 - (\bar C_1 + \frac{\bar C_2}{3}) ( 8G(m_c) + \frac43)\\
&& \quad -
\bar C_3 ( 8 G(m_c) - \frac43 G(0) - \frac{16}{3} G(m_b) + \frac{2}{27} ) \nonumber \\
&& \quad + \bar C_4 ( 4 G(0) - \frac83 G(m_c) + \frac{16}{3} G(m_b) + \frac{14}{9} ) \nonumber \\
&& - \bar C_5 (8 G(m_c) - 4 G(m_b) - \frac{14}{27}) \nonumber \\ 
&& - \bar C_6 ( \frac83 G(m_c) - \frac43 G(m_b) + \frac29) \nonumber \\
&& - \frac{\alpha_s}{4\pi} [ 2\bar C_1 (F_1^{(9)}(q^2) + \frac{F_2^{(9)}(q^2)}{6})
 + \bar C_2 F_2^{(9)}(q^2) + C_8^{\rm eff} F_8^{(9)}] \nonumber 
\end{eqnarray}
and
\begin{eqnarray}
&& C_7^{\rm eff} = C_7 - \frac49 \bar C_3 - \frac43 \bar C_4 
+ \frac19 \bar C_5 + \frac13 \bar C_6 \\
&& \quad - \frac{\alpha_s}{4\pi}[\bar C_2 F_2^{(7)}(q^2) + C^{\rm eff}_8 
F_8^{(7)}(q^2) ]\nonumber
\end{eqnarray}
They are expressed in terms of the modified Wilson coefficients $\bar C_{1-6}$,
which are defined by expressing the operators $O_{1-6}$ of Ref.~\cite{CMM}
in terms of the basis of \cite{BBL} using 4-dimensional Fierz identities.
They are given by \cite{fbcor1}
\begin{eqnarray}
&& \bar C_1 = \frac12 C_1\,,\qquad \bar C_2 = C_2 - \frac16 C_1 \\
&&\bar C_3 = C_3 - \frac16 C_4 + 16 C_5 - \frac83 C_6 \,,\quad 
\bar C_4 = \frac14 C_4 + 8 C_6 \nonumber \\
&& \bar C_5 = C_3 - \frac16 C_4 + 4 C_5 - \frac23 C_6\,,\quad 
\bar C_6 = \frac12 C_4 + 2 C_6 \nonumber 
\end{eqnarray}
where $C_i$ are the Wilson coefficients in the operator basis of Ref.~\cite{CMM}.
The $\bar C_i$ coefficients coincide with the Wilson coefficients in the basis of
Ref.~\cite{BBL}, but are different beyond leading log
approximation. The relation between the two sets of coefficients can be found in
Refs.~\cite{CMM,fbcor1}.
The effective Wilson coefficient $C_8^{\rm eff}$ is given by
\begin{eqnarray}
C_8^{\rm eff} = C_8 + \frac43 \bar C_3 - \frac13 \bar C_5\,.
\end{eqnarray}

The one-loop function $G(m_q)$ is given by
\begin{eqnarray}
G(m_q) = \int_0^1 \mbox{d} x x(1-x) \log\Big(
\frac{-q^2 x(1-x) + m_q^2 - i\epsilon}{\mu^2} \Big) \nonumber
\end{eqnarray}
The functions $F_{1,2}^{(9)}(q^2), F_2^{(7)}(q^2)$ appearing in the
2-loop matching conditions are listed in 
Eqs.~(54)-(56) of the second reference in Ref.~\cite{Asatrian}. The functions $F_8^{(7,9)}(q^2)$ are
given in Eqs.~(82), (83) of Ref.~\cite{fbcor1}.

We list here the functions $f_v$ and $\tau$ appearing in the expression of the
Wilson coefficient $c_1^{(V)}$ Eq.~(\ref{c1V})
\begin{eqnarray}
&& f_v(\omega,\mu) = \frac12 \log^2\frac{m_b^2}{\mu^2} - \frac52 \log\frac{m_b^2}{\mu^2} + 2\log \frac{m_b^2}{\mu^2}
\log\frac{\omega}{m_b} \nn\\
&& + 
2\log^2 \frac{\omega}{m_b} + 2 \mbox{Li}_2(1-\frac{\omega}{m_b}) + \frac{3\omega-2m_b}{m_b-\omega}
\log\frac{\omega}{m_b}\nn\\
&& + \frac{\pi^2}{12} + 6\,, \\
&& \tau(\omega, \mu)=\frac{\alpha_s C_F}{4\pi}
\Big[ \frac{\omega}{m_b-\omega}\log\frac{\omega}{m_b}  + \log\frac{m_b^2}{\mu^2} \Big] \,.
\end{eqnarray}

In the numerical evaluation of the Wilson coefficient $c_1^{(V)}$ we replace the
$\overline{MS}$ mass $m_b(\mu)$ with the pole mass $m_b^{\rm pole}$ using the one-loop
result
\begin{eqnarray}
m_b(\mu) = m_b^{\rm pole} \Big( 1 + \frac{\alpha_s C_F}{4\pi} (-6\log \frac{\mu}{m_b} - 4 )\Big)
\end{eqnarray}
and keeping only the term linear in $\alpha_s$.

Finally, we give here the function $t_\perp(x,m_c)$ appearing in the
Wilson coefficient of the subleading $O(\lambda)$ SCET$_{\rm I}$ operators.
This is given in Eq.~(27) of Ref.~\cite{fbcor1}, which we reproduce here for completeness
\begin{widetext}
\begin{eqnarray}\label{tperp}
&& t_\perp(x,m_c) = \frac{4m_B}{\bar x \omega} I_1(m_c) +
\frac{4q^2}{\bar x^2 \omega^2} [B_0(\bar x m_B^2 + x q^2, m_c) -
B_0(q^2, m_c)] 
\end{eqnarray}
with 
\begin{eqnarray}
&& B_0(s, m_c) = -2\sqrt{\frac{4m_c^2}{s}-1} \arctan 
\frac{1}{\sqrt{\frac{4m_c^2}{s}-1}}\\
&& I_1(m_c) = 1 + \frac{2m_c^2}{\bar x(m_B^2-q^2)} [L_1(x_+) + L_1(x_-) -
L_1(y_+) - L_1(y_-)] \nn
\end{eqnarray}

\end{widetext}
The function $L_1(x)$ and its arguments are defined as
\begin{eqnarray}
&& L_1(x) = \log\frac{x-1}{x} \log(1-x) - \frac{\pi^2}{6} + 
\mbox{Li}_2(\frac{x}{x-1}) \nn \\
&& x_\pm = \frac12 \pm \sqrt{\frac14 - \frac{m_c^2}{\bar x m_B^2 + x q^2}} \\
&& y_\pm = \frac12 \pm \sqrt{\frac14 - \frac{m_c^2}{q^2}}
\end{eqnarray}

\vspace{0.5cm}
\begin{acknowledgments}
D.P. would like to thank G. Hiller for useful discussions. 
We are grateful to Tim Gershon and Masashi Hazumi for comments on the manuscript.
This work was supported in part by the DOE under grant DE-FG03-97ER40546 (BG),
and by the DOE under cooperative research agreement
DOE-FC02-94ER40818 and by the NSF under grant PHY-9970781 (DP).
\end{acknowledgments}
\vfill

\end{document}